\documentclass[11pt]{article}
\usepackage{psfig}

\begin{document}

\section*{\centering THE SIZE-FREQUENCY DISTRIBUTION OF THE ZODIACAL CLOUD:
EVIDENCE FROM THE SOLAR SYSTEM DUST BANDS}
\begin{center} 
Keith Grogan, Stanley F. Dermott$^{*}$ and Daniel D. Durda$^{\dagger}$\\
NASA Goddard Space Flight Center, Code 681, Greenbelt, MD 20771 \\
\vspace{12pt}
$^{*}$Department of Astronomy, University of Florida, Gainesville, 
FL 32611 \\
$^{\dagger}$Southwest Research Inst., 1050 Walnut St., $\#$426, Boulder,
CO 80302 \\
\vspace{12pt}
Phone: (301) 286-4533 Fax: (301) 286-1752 \\
e-mail: grogan@stis.gsfc.nasa.gov \\
\vspace{2cm}
Submitted to {\it Icarus}, May 4 2000
\vspace{4cm}
\begin{itemize}
 \item Total number of pages: 
 \item Number of Figures:   
 \item Number of Tables:   
\item Key words: ASTEROIDS, DYNAMICS; INFRARED OBSERVATIONS; INTERPLANETARY
DUST; ZODIACAL LIGHT          
\end{itemize}
\end{center}
\clearpage
\begin{tabbing}
Proposed \=running header: \\
\>  \\
\> SIZE FREQUENCY DISTRIBUTION OF THE ZODIACAL CLOUD
\end{tabbing}
\begin{tabbing}
Correspon\=dence and proofs should be sent to: \\ 
\>  \\
\> Keith Grogan \\
\> NASA Goddard Space Flight Center \\
\> Code 681 \\
\> Blg. 21, Room 048 \\
\> Greenbelt, MD 20771, USA \\
\end{tabbing}

\clearpage

\begin{abstract}
Recent observations of the size-frequency distribution of zodiacal cloud 
particles obtained from the cratering record on the LDEF satellite (Love and 
Brownlee 1993) reveal a significant large particle population (100 micron 
diameter or greater) near 1 AU. Our previous modeling of the Solar System dust 
bands (Grogan et al 1997), features of the zodiacal cloud associated with the 
comminution of Hirayama family asteroids, has been limited by the fact that 
only small particles (25 micron diameter or smaller) have been considered. 
This was due to the prohibitively large amount of computing power required to 
numerically analyze the dynamics of larger particles. The recent availability 
of cheap, fast processors has finally made this work possible. Models of the 
dust bands are created, built from individual dust particle orbits, taking 
into account a size-frequency distribution of the material and the 
dynamical history of the constituent particles. These models are able to 
match both the shapes and amplitudes of the dust band structures observed by 
IRAS in multiple wavebands. The size-frequency index, $q$, that best matches 
the observations is approximately 1.4, consistent with the LDEF results in that 
large particles are shown to dominate. However, in order to successfully model 
the `ten degree' band, which is usually associated with collisional activity 
within the Eos family, we find that the mean proper inclination of the 
dust particle orbits has to be approximately 9.35$^{\circ}$, significantly 
different to the mean proper inclination of the Eos family (10.08$^{\circ}$). 
This suggests that either the ten degree band is produced from collisional 
activity near the inner edge of the family or that the inclinations of dust
particle orbits from the Eos family as a whole no longer trace the inclinations
of their parent bodies but have been degraded since their production.
\end{abstract}

\clearpage

\section{Introduction}

A little over fifteen years ago, the phenomenon of the zodiacal light was 
attributed to a smooth, lenticular distribution of cometary debris, centered 
on the Sun, lying in the plane of the ecliptic (see Giese et. al 1986 for a 
review). However, the launch of IRAS in 1983 revolutionized our knowledge of 
the interplanetary medium. For the first time, brightness profiles of the 
zodiacal cloud became available which clearly showed a level of structure, 
particularly near the ecliptic, which could not be explained by the previous 
paradigm. Figure 1 shows such a brightness profile of the zodiacal cloud, 
along with the results of passing the profile through a fast Fourier filter 
to isolate the near-ecliptic features. These features appear as `shoulders' 
superimposed on the background emission at roughly $\pm 10^{\circ}$, and a 
`cap' near the ecliptic plane. In the discovery paper, Low et. al (1984) 
suggest that these dust bands are traces of collisional debris within the 
main asteroid belt, based on a determination of their color temperature. 
This is an important point: the traditional source of the interplanetary 
dust complex was assumed to be the debris of short period comets (Whipple 
1967; Dohnanyi 1976). Although asteroid collisions should inject at least 
some material into the cloud, the lack of observational constraints had 
otherwise made the contribution of asteroidal material next to impossible to 
estimate.

A dust band is a toroidal distribution of dust particles with common 
inclinations. The dust particles themselves are asteroidal collisional debris.
Particles in cometary type orbits have high orbital eccentricities; planetary
gravitational perturbations produce large variations in these eccentricities
and these variations are coupled to those in the inclinations (Liou et al.
1995). Therefore even if a group of cometary type orbits initally had identical
inclinations, planetary perturbations would disperse those inclinations over a
wide range on a timescale of a few precession periods, showing that it is
impossible for a comet to produce a well defined dust band. 

A given asteroid undergoing a collision will break up producing debris
according to some size-frequency distribution. This distribution can be defined 
by the equation,
\begin{equation}
n(D) \sim D^{2-3q},
\end{equation}
where $D$ is the diameter of the particle. For a system in collisional 
equilibrium, q=11/6 (Dohnanyi 1969) and the distribution is dominated by 
small particles. Assuming the excess velocities after
escape are small compared with the mean orbital speed of an asteroid (15-20
km/s), the orbits of individual fragments will be similar as their orbital
elements will be only slightly perturbed from those of the parent asteroid
(Davis et al. 1979). Even a small initial distribution in relative velocity
(10-100 m/s), corresponding to a minor dispersion in semimajor axis $\Delta 
a/a$ (0.1-1\%) rapidly produces a ring of material over the parent asteroid's 
orbit (10$^{2}$-10$^{3}$ years). Secular precession acts upon the particles'
longitude of ascending node due to the effect of Jovian perturbations. To first
order,
\begin{equation}
\langle \dot{\Omega} \rangle = - \frac{3GM_{J}a^{3/2}}{4R_{J}^{3} 
\sqrt{GM_{\odot}}} \left[ 1+\frac{15}{8}\frac{a^{2}}{R_{J}^{2}} \right]
\end{equation}
where $\Omega$ is the longitude of node, $M_{J}$ is Jupiter's mass, $M_{\odot}$
is the Sun's mass, $R_{J}$ is the mean orbital distance of Jupiter (5.2 AU),
$a$ is the semimajor axis of a given particle and $G$ is the gravitational 
constant
(Sykes and Greenberg 1986). The rate of nodal regression is found from the
derivative of the above equation,
\begin{equation}
\Delta \langle \dot{\Omega} \rangle = \frac{9GM_{J}a^{3/2}}{8R_{J}^{3} 
\sqrt{GM_{\odot}}} \left[ 1+\frac{35}{8}\frac{a^{2}}{R_{J}^{2}} \right]
\frac{\Delta a}{a}.
\end{equation}
The time taken to distribute the nodes around the ecliptic to form a dust band
is then given by
\begin{equation}
\Delta t = \frac{2\pi}{\Delta\langle \dot{\Omega} \rangle}.
\end{equation}
For a collisional event at 2.2 AU with ejection velocities of 100 m/s, a dust
band would form after approximately 2 x 10$^{6}$ years. Now since particles in
inclined orbits spend a disproportionate amount of time at the extremes of
their vertical harmonic oscillations, a set of such orbits with randomly
distributed nodes will give rise to two apparent bands of particles
symmetrically placed above and below the mean plane of the system (Neugebauer
et al. 1984). This gives a natural explanation for the `shoulders' on the IRAS
profiles at approximately $\pm 10^{\circ}$. Similarly, the central `cap' may be
simply explained as a low inclination dust band. Any dispersion in the proper
inclinations of the dust particles will lead to the dust band profile 
appearing broader, with the peak intensity shifted to a lower latitude (Dermott
et al. 1990, Grogan et al. 1997).

A point of debate in the literature rests on whether the dust bands are
equilibrium or non-equilibrium features. In other words, are the dust bands
produced by a gradual grinding down of asteroid family members, or do they
represent regions of random, catastrophic disruptions in the asteroid belt? The
equilibrium model, first discussed by Dermott et al. (1984) and most recently
by Grogan et al. (1997), observes that the positions of the dust bands follow
the locations of the major Hirayama asteroid families. This would be the
natural consequence if the local volume density of dust, produced from
continual asteroid erosion, followed the local volume density of asteroids. The
catastrophic model follows from a discussion of dust band production rates
(assuming the random disruption of a small single asteroid of approximately
15km diameter) and dust band lifetimes (material will be removed by
Poynting-Robertson (P-R) drag). Following this logic Sykes and Greenberg (1986)
conclude that several dust bands should be visible at any given time. This is
in agreement with the IRAS observations and represents the main argument for
the non-equilibrium model. The question is an important one to answer, and has
implications for the investigation of the long-term evolution of the asteroid
belt. If the equilibrium model proves correct, then the dust bands can be used
as probes of collisional activity within their corresponding families and
ultimately employed to estimate the percentage contribution of asteroidal
material to the zodiacal dust complex. If the catastrophic paradigm is correct,
then individual dust band features cannot be related to given asteroids in the
belt with any confidence, and the question of the asteroidal contribution to
the cloud will be much more difficult to unravel.

\section{IRAS Observations of the Dust Bands}

Dust band structures are not observed independently from the rest of the 
zodiacal cloud. The IRAS observations consist of a series of line of sight
brightness profiles taken through the zodiacal cloud as a whole and to study 
the bands they must somehow be isolated from the remainder of the cloud. 
Various techniques have been employed in the literature for this purpose. 
Sykes (1990) uses a boxcar averaging method: this process averages data values 
over a given filter width (latitude bin) and subtracts that average from the 
central sample value. The filter is then shifted by one sample and the process 
repeated. This has the effect of smoothing the data, and the difference 
between the original data and the smoothed data gives the residuals which
are then associated with the dust bands. Reach (1992) and Jones and 
Rowan-Robinson (1993) assume some empirical form for the background 
component, and subtract this from the observations to produce residuals which
can then be associated with the dust bands. Finally, Fourier analysis has
been employed (Dermott et al. 1986, Sykes 1988, Grogan et al. 1997, 
Reach et al. 1997) where a smooth low frequency background is separated from 
the high frequency dust band residuals.

In this paper, the dust band residuals will be obtained by means of a fast 
Fourier filter. This filter is sampled at an equal number of points as the 
number of data points in the brightness profile. The frequency cut-off is 
defined by a simple coefficient $c$, a number which varies from 0 to 1 to 
represent the fraction of frequency points to remain after the high frequencies
are stripped from the Fourier transform. In other words, defining this 
coefficient equal to 1 would leave the complete Fourier transform intact. 
Figure 2 demonstrates how more and more of the original profile is incorporated
into the low frequency background as the constant $c$ increases. This is a 
dramatic illustration of the arbitrary nature of any filtering process and the 
danger of assuming the resultant residuals to represent the complete dust band 
structure.

The viewing geometry of the IRAS spacecraft was ideal for the study of the 
zodiacal cloud. The Medium Resolution (2' in scan) Zodiacal Observational 
History File (ZOHF) consists of 5757 sky brightness profiles, each providing 
a detailed view of the pole-to-pole cloud structure in a given line of sight 
defined by the ecliptic longitude of Earth, with most scans being taken at 
around 90$^{\circ}$ solar elongation. Towards the end of the survey the 
satellite covered elongation angles between 60$^{\circ}$ and 120$^{\circ}$, 
but most of these observations were contaminated by Galactic emission as the 
Galactic plane was at this point close to the ecliptic. The changes in shape 
and amplitude of the dust band residuals from profile to profile are caused by 
a combination of the complex three-dimensional structure of the dust bands 
themselves and also the observing geometry of the IRAS satellite.

The two primary causes for a change in the line of sight are (1) the longitude 
of Earth and (2) the solar elongation angle. The changes due to these two 
parameters are taken to be independent to first order, allowing a quantitative 
parameter to be associated with each. Changes in elongation angle produce a 
parallax effect: there is a change in the effective distance to the bands, and 
for small changes in elongation angle the effect can be assumed to be linear. 
The slope $d\gamma/d\epsilon$ for the change in peak latitude of the north or 
south dust band $\beta_{N}$ or $\beta_{S}$ with elongation angle can be found 
from a number of scans of a given longitude of Earth with varying solar 
elongation and this used to normalize the peak latitude that would be observed 
at a solar elongation of 90$^{\circ}$ (an example is shown in Figure 3 for a 
longitude of Earth of 227.3$^{\circ}$, trailing direction). Once this has been 
done, the normalized 
values of $\beta_{N}$ and $\beta_{S}$ may be used to obtain $<\beta>$, the 
mean north/south peak latitude, which may be plotted as a function of ecliptic 
longitude of Earth. This is shown in Figure 4 for the ten degree band in the 
25 $\mu m$ waveband. The sinusoidal variation indicates that the plane of 
symmetry of the bands, the plane about which on average the proper inclinations 
of the particles precess, is inclined to the ecliptic. This tilt of the plane 
of symmetry is due to the secular perturbations of the planets, and its 
orientation depends on the forced elements imposed on the dust particles. When 
viewed from Earth such a plane would appear as a sine curve, its amplitude 
equal to the inclination of the plane. Also, the displacement from the ecliptic 
will be equal in the trailing and leading directions at the ascending and 
descending nodes. Profiles of different longitudes of Earth can now be coadded 
using the parameters of the sine curve representing the peak latitude of the
bands due to their plane of symmetry being inclined to the ecliptic; this 
effect translates into a lateral shift that can be positive or negative 
depending on the longitude of Earth, and a minimum when Earth is at the 
forced nodes.

Individual IRAS scans were Fourier filtered and the dust band residuals coadded
in the above manner to produce several representative profiles around the sky 
normalized to a solar elongation angle of 90$^{\circ}$ with noise levels an 
order of magnitude less than the original scans. The results of this process 
for the 12, 25 and 60 $\mu m$ wavebands, leading and trailing, are shown in 
Figures 5-7. The dust band emission peaks in the 25 $\mu m$ waveband, although 
the bands are still clearly visible at 12 and 60 $\mu m$. The dust bands have a
lower amplitude but similar shape at 12 $\mu m$ compared to 25 $\mu m$, whereas
at 60 $\mu m$ the central band is less prominent with respect to the ten degree
band, an effect which is largely due to the filtering process. These 
observations contain a wealth of information about the structure of the dust 
bands; certain aspects, however, deserve special mention. Firstly, there exists 
a sinusoidal variation in the latitudes of peak brightness of the north and 
south ten degree band pair around the sky. This is due to the forced 
inclinations imposed on the dust band particle orbits by planetary 
gravitational perturbations as described above. Secondly there is a clear split 
in the central band and the amplitudes of each peak vary around the sky. The 
amplitudes of the north and south ten degree band pair also undergo such a 
variation except that this variation seems to be out of phase with the 
variation seen in the corresponding north and south peaks of the central band. 
The complex structure revealed by these observations underlines the point that
empirical models which attempt the describe the zodiacal cloud as a whole will 
always fall short in accounting for features such as these, and the problem 
demands a detailed dynamical treatment.

\section{A Physical Model for the Dust Bands}

The legacy of the IRAS, COBE and ULYSSES spacecraft is a realization that the
zodiacal cloud may consist of five distinct and significant components. These
are (1) the asteroidal dust bands (Dermott et al. 1984; Sykes and Greenberg
1986; Reach 1992; Grogan et al. 1997), (2) dust associated with other 
background (non-family) asteroids, (Dermott et al. 1994a) (3) dust associated 
with cometary 
debris (Sykes and Walker 1992; Liou and Dermott 1995), (4) the Earth's resonant
ring (Dermott et al 1994b), and (5) interstellar dust (Grun et al. 1994; Grogan
et al. 1996). It is also possible that a significant proportion of
interplanetary dust particles originate in the Kuiper belt (Flynn 1996, Liou
et al. 1996).
The approach of the Florida group (Dermott et. al) has been to
place constraints on the origin and evolution of material of a given source
from both dynamical considerations and observational data. Given a postulated
source of particles, the aim is to describe (1) the orbital evolution of these
particles, including P-R drag, using equations of motion that include the solar
wind, light pressure and planetary gravitational perturbations, and (2) the
thermal and optical properties of the particles and their variation with
particle size. Once the dust particles and their distribution have been
specified along these lines, a line-of-sight integrator is employed to not only
view the model cloud but to reproduce the exact viewing geometry of any
particular telescope in any given waveband. The result is a series of model
profiles which can then be compared with observations.

Amongst the various forces acting upon the dust particles the most obvious is
solar gravity,
\begin{equation}
F_{grav}(r)=G\frac{Mm}{r^{2}}
\end{equation}
where $G$ is the gravitational constant and $M$ is the solar mass. Scattering
and absorption of solar radiation by a dust particle involve the transfer of
momentum and hence to a radiation pressure directed radially outwards (Burns et
al. 1979). For spherical particles radiation pressure takes the value
\begin{equation}
F_{rad}(r)=\frac{SA}{c}Q_{pr}
\end{equation}
where $S = L/4\pi r^{2}$ is the radiation flux density at distance $r$, $L$ is
the solar luminosity and $Q_{pr}$ is an efficiency factor averaged over the
solar spectrum which can be calculated using, for example, Mie theory (Bohren
and Huffman, 1983). Radiation pressure is usually expressed as the ratio of its
strength to the gravitational attraction, which for spherical particles is 
given by
\begin{equation}
\beta=\frac{F_{rad}}{F_{grav}}=5.7 \times 10^{-5} \frac{Q_{pr}}{\rho s}
\end{equation}
where $s$ is the particle radius and $\rho$ and $s$ are given in cgs units.
Roughly speaking, radiation pressure balances gravity for a 1 $\mu$m particle.
The component of radiation pressure tangential to the particle orbit gives
rise to the phenomenon known as Poynting-Robertson (P-R) drag, which results
in an evolutionary decrease in both the semi-major axis and eccentricity
of the particle orbit. These changes in the orbital elements can be given by
\begin{equation}
\frac{da}{dt}=-\alpha\frac{2+3e^{2}}{a(1-e^{2})^{3/2}}
\end{equation}
\begin{equation}
\frac{de}{dt}=-\frac{5\alpha e}{2a^{2}(1-e^{2})^{1/2}}
\end{equation}
\begin{equation}
\frac{di}{dt}=0
\end{equation}
where
\begin{equation}
\alpha=\frac{3.35 \times 10^{-10} Q_{pr}}{s(m)}AU^{2}/yr
\end{equation}
(Wyatt and Whipple 1950). The consequence is that the orbit shrinks and
circularizes, and a particle in a circular orbit at heliocentric distance $r$
spirals into the Sun in a time 
\begin{equation}
\tau_{pr}=700 \: s(\mu m) \: \rho (g/cm^{3}) \: r^{2} (AU) \: \langle Q_{pr} 
\rangle \: yrs
\end{equation}
This equates to several $10^{4}$ years for a `typical' particle (10 $\mu$m,
2.5 g/cm$^{3}$, initial $r$=2 AU).

Now consider the motion of dust particles under the effects of planetary
gravitational perturbations. When the eccentricity and inclination are small,
the solutions of the Lagrangian equations of motion for the eccentricity and
pericenter variations may be completely decoupled from the inclination and node
variations. These pairs of elements have simple vectorial representations and
may be decomposed into components known as the proper elements and the forced
elements of the orbit. The proper elements represent the stable long-term
averages that remain after removal of planetary perturbations. The variations
due to these perturbations are the forced elements, which can themselves be
separated into three categories: (1) secular (long period) perturbations; (2)
resonant (short period) perturbations; (3) transient (scattering)
perturbations. These perturbations acting on a small body in orbit about the
Sun precess the node and pericenter and over sufficiently long intervals the
distributions of these elements become essentially random. Figure 8 shows a
schematic of the vectorial relationship between the total (osculating) elements
($I$, $\Omega$), the proper elements ($I_{p}$, $\Omega_{p}$) and the forced 
elements ($I_{f}$, $\Omega_{f}$) in ($I$ cos $\Omega$, $I$ sin $\Omega$) space. 
The distribution 
is displaced from the origin due to the forced elements and the radius of the
distribution represents the proper elements. An equivalent relationship
exists for eccentricity and pericenter. Figures 9 and 10 show the evolution
of 249 Koronis dust particles migrating from the asteroid belt toward the Sun
(Kortenkamp and Dermott 1998). Secular perturbations, primarily from Jupiter
and Saturn, vary with time and semi-major axis and act to change the forced
elements of the distribution as the wave migrates into the inner Solar System.
The orbital eccentricities also decay due to P-R drag and solar wind drag.

Dermott et al. (1984) were the first to suggest that the Solar System dust
bands may originate in the three prominent Hirayama asteroid families (Eos,
Themis and Koronis). To confirm their hypothesis of the asteroidal origin of
the dust bands, and to facilitate the investigation of the zodiacal cloud in
general, SIMUL, a three-dimensional numerical model was constructed (Dermott 
et al. 1988).
The basic ideas and assumptions behind SIMUL are as follows.
\begin{enumerate}
\item A cloud is represented by a large number of dust particle orbits. The
total cross-sectional area of the cloud is divided equally among all the
orbits.
\item The orbital elements of the dust particle orbits in the cloud can be
decomposed into proper and forced vectorial components. When inclination and
eccentricity are low, as is typically the case for asteroidal type orbits, at
any given time the
forced elements are independent of the proper elements and depend only on the
semimajor axis and the particle size.
\item As a first approximation, the dust particles in the cloud produced by 
asteroid families have the
same mean proper elements as those of the parent bodies, although the Gaussian
distribution of these elements is a free parameter.
\item The forced elements as a function of semimajor axis are calculated using
secular perturbation theory via direct numerical integrations, as outlined 
above.
\item Along each of the orbits, particles are distributed according to Kepler's
Law. Once the spatial distribution of the orbits is specified, space is divided
into a sufficiently large number of ordered cells and then every orbit is
investigated for all the possible cross-sectional area contributions to each of
the space cells. The model generates a large three-dimensional array which
serves to describe the spatial distribution of the effective cross-sectional
area.
\item The viewing geometry of any telescope can be reproduced exactly by
calculating the Sun-Earth distance and ecliptic longitude of Earth at the
observing time and setting up appropriate coordinate systems. In this way,
IRAS-type brightness profiles can be created and compared with the observed
profiles.
\end{enumerate}

In order to compare the results of the SIMUL modeling algorithm with the IRAS
observations, the filtering problem - the fact that a substantial percentage 
of the dust band signal is indistinguishable from the background - must be 
addressed. An iterative process (Dermott et al. 1994a, Grogan et al. 1997) is 
used to determine the low-frequency component of the dust band and therefore 
bypass this filtering problem. Figure 11 shows how this is achieved. Panel (a) 
shows a raw model dust band having the same viewing geometry as an observed 
background, produced by filtering off the high-frequency dust band component. 
In the first iteration (a) is added to (b) and the sum is filtered to obtain 
(c), a filtered model dust band (smooth curve) - the observed dust bands 
(noisy curve) are also plotted for comparison. The background obtained from 
this iteration, shown in panel (d) is of a higher intensity than the original 
background due to the fact that it contains two low-frequency dust band 
components, one from the addition of the model dust band and one from the 
actual dust band in the original observed background (the high frequency 
component of which was removed in the creation of the observed background). 
In other words, the difference between these two backgrounds gives the extent 
of the low-frequency dust band component. In the final iteration we subtract 
the excess intensity shown in panel (d) from the original background (b) and 
add (a) before filtering to obtain the final dust band model (e) and the final 
background (f) that agree with the observations. Thus, by using the same 
filter in the modeling process that we use to define the observed dust bands, 
and iterating, we are able to bypass the arbitrary divide associated with the 
filter.

\section{Results}

This work differs from our previous modeling of the dust bands (Grogan et al. 
1997) in that our models include a size-frequency distribution, rather than 
being composed of particles of a single size. This is critical in our efforts 
to provide a model of the dust bands that can match the IRAS observations in 
multiple wavebands. Particles ranging in size from 1 to 100 $\mu m$ are 
included, each of which are assumed to be Mie spheres composed of astronomical 
silicate (Draine and Lee 1984). The lower end cut-off is determined by the fact 
that contribution to the thermal emission from particles smaller than this
size is negligible. The upper cut-off follows from the fact that in the 
zodiacal cloud, the P-R drag lifetime is comparable to the collisional 
lifetime for a particle of about this size (Leinert and Gr\"{u}n 1990). 
However, the inclusion of a wide range of particle sizes can only be achieved 
with an understanding of their dynamical history, so that their orbital 
distributions can be properly described in the SIMUL algorithm. This is 
achieved using the RADAU fifteenth order integrator program with variable 
time steps taken at Gauss-Radau spacing (Everhart 1985), with which we perform 
direct numerical integration of the full equations of motion of interplanetary 
dust particles (IDPs) of various sizes. Our simulations include seven planets 
(Mercury and Pluto excluded) and account for both P-R drag and solar wind drag. 
The average force due to the solar wind drag is taken to be 30\% of the P-R 
drag force, varying with the 11-year solar cycle from 20\% to 40\% (Gustafson 
1994). In this way we are able to build a description of both the proper and 
forced elements of the particles and their variation with heliocentric 
distance from their simple vectorial relationship shown in Figure 8. Because
the forced elements vary as a function not only of semi-major axis but also
of time, each wave of particles (as shown in Figures 9 and 10) is started
at different times in the past, such that when the waves reach the present 
they span the full range of semi-major axis from the asteroid belt into the
Sun. In this way a snapshot of the present day forced element distribution
is constructed. Figure 
12 shows the variation with heliocentric distance of the forced inclination 
(top) and forced longitude of ascending node (bottom) of 4, 9, 14, 25 and 
100 $\mu m$ diameter IDPs in the zodiacal cloud. As the particle size 
increases, its P-R drag lifetime increases and it therefore spends longer in 
secular resonances near the inner edge of the asteroid belt. This causes the 
forced inclination of a 100 micron 
diameter particle to approach 6$^{\circ}$ interior to 2 AU. An equivalent
diagram for the forced eccentricity and forced pericenter is shown in Figure
13.

Dust band models are produced via SIMUL in the following manner.
\begin{enumerate}
\item A model to account for the central band is created by using two 
distributions of orbits having mean semi-major axis, proper eccentricity, 
proper inclination and dispersions equal to those found in the Themis and 
Koronis families. The proper elements found from the numerical integrations 
are added vectorially to find the osculating orbital elements, and the material 
is distributed into the inner Solar System as far as 2 AU according to P-R 
drag (a $1/r^{\gamma}$, $\gamma=1.0$ distribution). The size-frequency 
distribution of material in 
the observed dust band is investigated by varying the size-frequency index $q$ 
of particles in the model.
\item A model to account for the ten degree band is created from Eos type 
orbits, in that their mean semi-major axis and eccentricity are equal to those 
found in the Eos families. However, in order to improve upon previous modeling 
of the ten degree band the mean proper inclination of the distribution was 
allowed to vary within the range of proper inclinations found in the Eos 
family. A best fit was found at a mean proper inclination of 9.35$^{\circ}$ 
with a dispersion of 1.5$^{\circ}$. Again, the proper elements are added 
vectorially, the material distributed into the inner Solar System as far as 
2 AU, and a size-frequency distribution applied.
\item We do not create a model for the zodiacal background but instead add the 
model dust band profiles to the observed background obtained from applying the 
Fourier filter to the corresponding raw IRAS observation. The total is then 
filtered using the iterative procedure described above so that the resultant 
model residual can be directly compared with the observed dust bands. For a 
given size-frequency index $q$, the 
total surface area of material associated with the model bands is adjusted 
until the amplitudes of the 25 $\mu m$ model dust bands matches the 25 $\mu m$ 
observations; $q$ can then be varied until a single model provides a match in 
amplitude to the 12, 25 and 60 $\mu m$ observations simultaneously.
\end{enumerate}

Figures 14-16 show the best results of our modeling, comparing the dust band 
observations (solid curves) to the dust band models (dotted curves) in the 12, 
25 and 60 $\mu m$ wavebands. The models were constructed as described above, 
and have a size-frequency index $q$ equal to 1.43. Large particles dominate 
this distribution. Table 1 lists the parameters used in the model components. 
The amplitudes in all wavebands are well matched, and the shapes of the dust 
band models describe the variation in shape of the observations around the sky 
very well. Figure 17 shows how the goodness of fit of our models changes as a 
function of size-frequency index $q$ for a single longitude of Earth.
This has been obtained for the ten degree band by calculating the root mean 
square (observation - model) over two five degree wide latitude bins to cover 
the north and south bands for both the 12 and 60 micron wavebands.
In essence, the wavebands act as filters through which different particle 
sizes in the cloud are seen. The 12 $\mu m$ waveband preferentially detects 
emission from the smaller particles, and the 60 $\mu m$ waveband preferentially 
detects emission from the larger particles. Therefore,
\begin{enumerate}
\item When $q$ is too high, too many small particles are included in the
model, and the ampitudes of the 12 $\mu m$ models are too large. In addition, 
too few large particles are included and the amplitudes of the 60 $\mu m $ 
models are too small. This effect can be seen in Figures 18-19, in which dust 
band models have been produced with $q$=1.83, appropriate for a system in 
collisional equilibrium. 
\item When $q$ is too low, too many large particles are included in the model. 
This leads to the distribution of forced inclinations in the model to be 
skewed too much towards the large end, and the model profiles are shifted in 
latitude with respect to the observations, degrading the fit. This effect is 
much smaller than the amplitude effect for high $q$, and will only be properly 
quantified when a fuller description of the action of large particles at the 
2 AU secular resonance has been produced.
\end{enumerate}

A clear result is that a high size-frequency index $q$, in which small 
particles dominate, fails to account for the observations of the Solar System 
dust bands. This index has to be reduced to the point where large particles 
dominate the distribution. This is consistent with the cratering record on 
the LDEF satellite, shown in Figure 20, which suggests a $q$ of approximately 
1.15 at Earth and a peak in the particle diameter at around 100-200 micron. 
Since the Fourier filter preferentially isolates material exterior to the 2 AU
secular resonance (in the inner Solar System the dust band material is
dispersed into the background cloud due to the action of secular resonances),
our results are more indicative of the size-frequency index of dust
in the asteroid belt. We do not mean to claim that the size-frequency
index $q$ is a constant throughout the main-belt: the true nature of the
distribution will be a complex function of dust production rates, P-R drag
rates, collisional lifetimes and the nature of particle-particle collisions.
Consequently, the size distribution will presumably be some function of
heliocentric distance. However, in describing in main belt region as a whole,
we do claim that large particles appear to dominate the dust band emission
over small particles.

One concern that must be addressed is the possibility that the observed 
relative amplitudes 
of the dust band material are driven by the optical properties of the dust
particles, and not the size-frequency index. For this reason, we have 
repeated the modeling process assuming the particles to be made of organic
refactory material (Li and Greenberg 1997). Figure 21 shows the variation 
with wavelength and particle diameter of the absorption efficiencies of both
astronomical silicate (top) and organic refractory material (bottom), 
calculated using Mie theory. One striking difference between the two, 
particularly relevant for this discussion, is that emission at longer
wavelengths for large particles is highly attenuated for the organic
refractory material compared to the astronomical silicates. Figure 22
shows how the residuals obtained in the modeling process are affected by
the change in the composition of the dust particle. The 12 micron residuals
strongly reinforce the result obtained with astronomical silicate that a
low size-frequency index $q$ is required to match the observations. The
evidence at 60 micron is less clear, where we are hindered by the low 
emissivity of this material at longer wavelengths for larger particles.
Even so, the residuals are decreasing with decreasing $q$. The consistent
picture is that the dust band distribution is dominated by particles at
the large end of the size range. 

\section{Discussion}

The results presented in this paper improve upon those reported in a
previous paper (Grogan et al. 1997), particularly in regard to the ten
degree band associated with the Eos family. In order for a dust band model 
to match the observations, it needs to fit both the latitude of peak flux 
(driven by the mean proper inclination of the particles) and the width of 
the dust band feature (a function of the dispersion in proper inclinations).
Previously, the dispersion in proper inclinations of the Eos dust particles 
was reported at a relatively high 2.5$^{\circ}$, which minimized the residuals
while the mean proper inclination of the particles was fixed at the mean
proper inclination of the Eos asteroid family. In this paper, smaller residuals 
are found when the mean proper inclination of the particles is allowed to float
as a free parameter; the best fit then corresponds to a mean proper
inclination of 9.35$^{\circ}$ and a dispersion of only 1.5$^{\circ}$.

Figure 23 shows the members of the Eos asteroid family in (e,i) space as 
determined by the hierarchical clustering method (Zappala et al. 1995). Shown 
on this diagram is the position of the mean proper inclination of the ten 
degree band model. The consequence is that the ten degree dust band material
is not tracing the orbital element space of the Eos family as a whole, as 
would perhaps be expected from the equilibrium model. Either the collisional
activity is occurring near the inner edge of the Eos family, or the 
inclinations of dust particle orbits originating from the Eos family as a 
whole no longer trace the inclinations of their parent bodies but have been 
degraded since their production. If some mechanism was degrading the dust
particle orbits it would presumably apply to particles from all sources, but 
may be more easily observed within the Eos family owing to its high inclination.
Trulsen and Wikan (1980) have suggested based on their numerical simulations
that the combined influence of P-R drag and collisions acts to decrease both 
the mean eccentricity and inclination of dust particle orbits. This subject
is however open to debate; the nature of collisions between interplanetary
dust particles is still poorly understood. 
Figure 24 shows the cumulative surface area as a function of 
different size-frequency distrubtion indices for the Eos, Themis and Koronis 
families and also a single 15km diameter 
asteroid. At first this appears to contradict our result that a low $q$ of 
around 1.43 is needed to model the dust bands. However, the diagram is set up 
such that size-frequency distribution is constant from the source point all 
the way down to the smallest IDPs, which we know is not the case since P-R 
drag will act to preferentially remove the small particles. In reality, the 
size-frequency distribution will change from the large to the small end of the 
distribution, and will also be a function of heliocentric distance. 
The diagram does suggest that for a single asteroid to be responsible for the 
ten degree dust band, the size-frequency index of the collisional debris would 
initially have needed to be extremely high to produce the surface area
required to match the observations. 

The justification of cutting off the distribution of dust band material at 2 AU 
is essentially given by Figure 12. As the particles move out of the asteroid 
belt the action of the secular resonance disperses them into the background 
cloud, an effect which is more marked as the particle size increases. For this 
reason the Fourier filter is particularly sensitive to material located in the 
asteroid belt, and models that confine the material to the asteroid belt match 
the observations very well. In the future, our models will populate the inner 
Solar System as well as the main-belt region, but to do this properly we will 
have to:
\begin{enumerate}
\item Investigate the dynamical history of a much greater number of particle 
sizes than the five sizes we have considered so far in order to properly 
account for their behavior at the 2 AU secular resonance;
\item Take into account collisional processes: larger particles will have 
shorter collisional lifetimes compared to their P-R drag lifetimes and will 
therefore not penetrate as far into the inner Solar System. Each distribution 
of orbits of a given particle size will therefore have a natural inner edge 
defined by the lifetime of the particles in the cloud.
\end{enumerate}
However, we can obtain an estimate for the dust band contribution to the 
zodaical cloud as a whole by simply extending our best fit dust band models to 
populate the inner Solar System. The distribution of orbits obtained in this 
manner will not be exactly correct, due to our insufficient treatment of the 
secular resonance, but will still be reasonably accurate in terms of the total 
surface area associated with the dust bands. Figure 25 compares the thermal 
emission obtained from this raw dust band model to the corresponding IRAS 
profile in the 25 $\mu m$ waveband. The result is shown for inner Solar System 
distributions of material corresponding to $\gamma$=1.0, as expected for a 
system evolved by P-R drag, and $\gamma$=1.3 as predicted in parametric models 
of the zodiacal cloud, most recently Kelsall et al. (1998). The dust bands 
appear to contribute 
approximately 30\% to the total thermal emission. Also shown is the amplitude 
of the dust band material confined
to the main belt (exterior to 2 AU), which represents the component of the
dust band material isolated by the fast Fourier filter. This clearly shows
the extent to which the dust band contribution is underestimated if it is
assumed that the filtered dust band observations represent the entirety of 
the dust band component of the cloud. Figure 26 shows the ratio of areas of 
material associated with the entire main belt asteroid population and all
families, for asteroid diameters greater than 1 km. The best fit lines have
a slope corresponding to a size-frequency index $q=1.795$.
This diagram can be 
used to estimate the total contribution of main belt asteroid collisions to 
the dust in the zodiacal cloud, by extrapolating the observed size 
distributions of larger asteroids in both populations assuming a collisional
equilibrium power law size distribution. The result is that the main belt
asteroid population contributes approximately three times the dust area of
the Hirayama families alone, and the total asteroidal contribution to the 
zodiacal cloud could account for almost the entireity of the interplanetary 
dust complex. In reality, evolved size distributions are more complex than
simple power laws (Durda et al. 1998) and the size distribution of
individual asteroid families likely preserve some signatures of the original
fragmentation events from which they were formed. However, small dust-size
particles and their immediate parent bodies have collisional lifetimes in 
the main belt that are considerably shorter than the age of the Solar System
or the major asteroid families. Thus the dust size distributions associated
with both the background main belt and family asteroids may well be
considered to have achieved an equilibrium state, with total areas related to
the equivalent volumes of the original source bodies in each population. 
An alternative, and perhaps more satisfactory, approach to 
obtaining the total asteroidal contribution to the zodiacal cloud will be to 
apply our methods to the main-belt asteroid population in the same way we have
investigated the dust bands. This is the subject of a future paper.

The origin of the large dispersion in proper inclination (1.5$^{\circ}$) 
required to successfully model the ten degree band, in rough agreement with 
the 1.4$^{\circ}$ found by Sykes (1990) and the  2$^{\circ}$ found by Reach 
et al. (1997), remains unclear. Dispersion in 
inclination due to the Lorentz force is expected to behave such that the root 
mean square of the dispersion will increase with the square root of the 
distance traveled, and will be inversely proportional to the cube of the 
radius of the particle (Leinert and Gr\"{u}n 1990). Morfill and Gr\"{u}n (1979) 
report a value of only 0.3$^{\circ}$ for a particle of 1 $\mu m$ radius by the 
time it has spiraled in to 1 AU from the asteroid belt after 3000 years,
with that expected for a 100 micron particle to be significantly less. 
Subsequent treatments by Consomagno (1979), Barge et al. (1982) and 
Wallis and Hassan (1985) differ by more than an order of magnitude due to the 
lack of detailed knowledge of the magnetic field structure. A more likely 
source of the dispersion is simply the action of the secular resonance at 2 AU. 
However, this leaves open the question of why a large dispersion is required 
to model the ten degree band, and only the small dispersion of the Themis and 
Koronis families is required to successfully reproduce the central band 
observations. One answer may be that the emission associated with the central 
band is due to relatively recent collisions within these families. Figure 27
shows the variation with time of the total cross-sectional area associated
with the main belt and describes the stochastic breakup of asteroidal fragments.
This numerical approach to describing the collisional evolution of the
asteroid belt is detailed by Durda and Dermott (1997).
The initial main belt mass is taken to be approximately three times greater
than the present mass (Durda et al. 1998); this population evolves after
4.5 Gyr to resemble the current main belt. The calculation
is performed for particles from 100 $\mu m$ through the largest asteroidal 
sizes, with a fragmentation index $q=1.90$. The dust 
production rate in the main asteroid belt becomes more stochastic with time
following a relatively smooth decrease in area as the small particles are
created directly from the breakup of the parent body are destroyed. The 
spikes in the dust production are due to the breakup of small to intermediate
size asteroids. Therefore while the observable volume of a family may decay at
a fairly constant and well-defined rate, the total area of dust associated with
the family during that time may fluctuate by an order of magnitude or more.

We have shown in this paper how the Solar System dust bands can be investigated 
and used as a tool for addressing fundamental questions about the nature of the
zodiacal cloud and the origin of the material from which it is composed. A key
component of this process has been the realization that large particles play a 
dominating role in the structure of the cloud and that their dynamical
histories need to be included in any physically motivated model. In the future
we will extend our knowledge of the dust dynamics to a wider range of particle 
sizes, and address the main-belt contribution as well as the dust band 
component on the way to our ultimate goal of providing a global model for
the zodiacal emission. 
 
\clearpage
\vspace{-0.5cm}
\section*{\centering{\normalsize References}}
\vspace{-0.2cm}

\begin{list}
{}{\setlength{\rightmargin}
{\leftmargin}}

\item{\sc Barge, P., Pellat, R. and Millet, I.} 1982. Diffusion of Keplarian 
motions by a stochastic force. II. Lorentz scattering of interplanetary dusts. 
{\it Astron. Astrophys.} {\bf 115}, 8-19.

\item{\sc Bohren, C.F. and Huffman, D.R.} 1983. {\it Absorption and Scattering
of Light by Small Particles}, Wiley, New York.

\item{\sc Burns, J.A., Lamy, P.L. and Soter, S.} 1979. Radiation Forces on
Small Particles in the Solar System. {\it Icarus} {\bf 40}, 1-48.

\item{\sc Consomagno, G.} 1979. Lorentz scattering of interplanetary dust.
{\it Icarus} {\bf 38}, 398-410. 

\item{\sc Davis, D.R., Chapman, C.R., Greenberg, R., Weidenschilling, S.J. and
Harris, A.W.} 1979. Collisional evolution of asteroids. In {\it Asteroids}, ed.
T. Gehrels, 528-557. Univ. of Arizona Press, Tuscon.

\item{\sc Dermott, S.F., Nicholson, P.D., Burns, J.A. and Houck, J.R.} 1984.
Origin of the Solar System dust bands discovered by IRAS. {\it Nature} {\bf
312}, 505-509.

\item{\sc Dermott, S.F., Nicholson, P.D. and Wolven, B.} 1986. Preliminary
Analysis of the IRAS Solar System Dust Data. In {\it Asteroids, Comets, Meteors
II}, eds. C.-I. Lagerkvist, B.A. Lindblad, H. Lundstedt and H. Rickman, 
583-594, Reprocentralen HSC, Uppsala.

\item{\sc Dermott, S.F., Nicholson, P.D., Kim, Y., Wolven, B. and Tedesco, E.F}
1988. The Impact of IRAS on Asteroidal Science. In {\it Comets to Cosmology},
ed. A. Lawrence, 3-18. Springer-Verlag, Berlin.

\item{\sc Dermott, S.F., Nicholson, P.D., Gomes, R.S. and Malhotra, R.} 1990.
Modeling the IRAS Solar System dust bands. {\it Adv. Space Res.} {\bf 10(3)},
171-180.

\item{\sc Dermott, S.F., Durda, D.D., Gustafson, B.\AA.S., Jayaraman, S., Liou,
J.C. and Xu, Y-L.} 1994a. Zodiacal Dust Bands. In {\it Asteroids, Comets, 
Meteors 1993}, eds. A. Milani, M. Martini and A. Cellino, 127-142. Kluwer, 
Dordrecht.

\item{\sc Dermott, S.F., Jayaraman, S., Xu, Y-L., Gustafson, B.\AA.S. and Liou,
J.C.} 1994b. A circumsolar ring of asteroidal dust in resonant lock with the
Earth. {\it Nature} {\bf 369}, 719-723. 

\item{\sc Dohnanyi, J.S.} 1969. Collisional model of asteroids and their debris.
{\it J. Geophys. Res.} {\bf 74}, 2531-2554.

\item{\sc Dohnanyi, J.S.} 1976. Sources of interplanetary dust, asteroids.
In {\it Interplanetary Dust and Zodiacal Light, Lecture Notes in Physics
Vol. 48}, eds. H. Els\"{a}sser and H. Fechtig, 187-205. Springer, Berlin.

\item{\sc Draine, B.T. and Lee, H.M.} 1984. Optical properties of interstellar
graphite and silicate grains. {\it Ap.J} {\bf 285}, 89-108.

\item{\sc Durda, D.D. and Dermott, S.F.} 1997. The Collisional Evolution of
the Asteroid Belt and Its Contribution to the Zodiacal Cloud. {\it Icarus}
{\bf 130}, 140-164.

\item{\sc Durda, D.D., Greenberg, R. and Jedicke, R.} 1998. Collisional
Models and Scaling Laws: A New Interpretation of the Shape of the Main-Belt
Asteroid Size Distribution. {\it Icarus} {\bf 135}, 431-440.

\item{\sc Everhart, E.} 1985. An efficient integrator that uses Gauss-Radau
spacings. In {\it Dynamics of Comets}, eds. A. Carusi and
G.B. Valsecchi, 185-202. Reidel, Dordrecht.

\item{\sc Flynn, G.J.} 1996. Collisions in the Kuiper Belt and the Production
of Interplanetary Dust Particles. {\it Meteoritics and Planet. Sci.} {\bf 31},
A45.

\item{\sc Giese, R.H., Kneissel, B. and Rittich, U.} 1986. Three-dimensional
zodiacal cloud, a comparative study. {\it Icarus} {\bf 68}, 395-411.

\item{\sc Grogan, K., Dermott, S.F. and Gustafson, B.\AA.S.} 1996. An
Estimation of the Interstellar Contribution to the Zodiacal Thermal Emission.
{\it Ap.J} {\bf 472}, 812-817.

\item{\sc Grogan, K., Dermott, S.F., Jayaraman, S. and Xu, Y-L.} 1997. Origin
of the ten degree dust bands. {\it Planet. Space Sci.} {\bf 45}, 1657-1665.

\item{\sc Gr\"{u}n, E., Gustafson, B.\AA.S., Mann, I., Baguhl, M., Morfill,
G.E., Staubach, P., Taylor, A. and Zook, H.A.} 1994. Interstellar Dust in the
Heliosphere. {\it Astron. Astrophys.} {\bf 286}, 915-924.

\item{\sc Gustafson, B.\AA.S.} 1994. Physics of zodiacal dust. {\it Ann. Rev.
Earth Planet. Sci.} {\bf 22}, 553-595.

\item{\sc Jones, M.H. and Rowan-Robinson, M.} 1993. A physical model for
the IRAS zodiacal dust bands. {\it MNRAS} {\bf 264}, 237-247.

\item{\sc Kelsall, T., Weiland, J.L., Franz, B.A., Reach, W.T., Arendt, R.G.,
Dwek, E., Freudenreich, H.T., Hauser, M.G., Moseley, S.H., Odegard, N.P., 
Silverberg, R.F. and Wright E.L.} 1998. The COBE Diffuse Infrared Background
Experiment Search for the Cosmic Infrared Background. II. Model of the
Interplanetary Dust Cloud. {\it Ap.J} {\bf 508}, 44-73.

\item{\sc Kortenkamp, S.J. and Dermott, S.F.} 1998. Accretion of Interplanetary
Dust Particles by the Earth. {\it Icarus} {\bf 135}, 469-495.

\item{\sc Leinert, C. and Gr\"{u}n, E.} 1990. Interplanetary Dust. In {\it 
Space and Solar Physics, Vol. 20, Physics and Chemistry in Space: Physics of 
the Inner Heliosphere I}, eds. R. Schween and E. Marsch, 207-275. Springer,
Berlin.

\item{\sc Li, A. and Greenberg, J.M.} 1997. A unified model of interstellar 
dust. {\it Astron. Astrophys.} {\bf 331}, 566-584.

\item{\sc Liou, J.C., Dermott, S.F. and Xu, Y-L.} 1995. The contribution of
cometary dust to the zodiacal cloud. {\it Planet. Space Sci.} {\bf 43},
717-722.

\item{\sc Liou, J.C., Zook, H.A. and Dermott, S.F.} 1996. Kuiper Belt Dust
Grains as a Source of Interplanetary Dust Particles. {\it Icarus} {\bf 124},
429-440.

\item{\sc Low, F.J., Beintema, D.A., Gautier, T.N., Gillet, F.C., Beichmann,
C.A., Neugebauer, G., Young, E., Aumann, H.H., Boggess, N., Emerson, J.P.,
Habing, H.J., Hauser, M.G., Houck, J.R., Rowan-Robinson, M., Soifer, B.T.,
Walker, R.G. and Wesselius, P.R.} 1984. Infrared Cirrus: New Components of
the Extended Infrared Emission. {\it Ap.J} {\bf 278}, L19-L22.

\item{\sc Morfill, G.E. and Gr\"{u}n, E.} 1979. The motion of charged dust 
particles in interplanetary space. I. The zodiacal dust cloud. {\it Planet. 
Space Sci.} {\bf 27}, 1269-1282.

\item{\sc Neugebauer, G., Beichman, C.A., Soifer, B.T., Aumann, H.H., Chester,
T.J., Gautier, T.N., Gillett, F.C., Hauser, M.G., Houck, J.R., Lonsdale, C.J.,
Low, F.J. and Young, E.T.} 1984. Early results from the infrared astronomical
satellite. {\it Science} {\bf 224}, 14-21.

\item{\sc Reach, W.T.} 1992. Zodiacal Emission III. Dust Near The Asteroid
Belt. {\it Ap.J} {\bf 392}, 289-299.

\item{\sc Reach, W.T., Franz, B.A. and Weiland, J.L.} 1997. The 
Three-Dimensional Structure of the Zodiacal Dust Bands. {\it Icarus} 
{\bf 127}, 461-484.

\item{\sc Sykes, M.V.} 1988. IRAS Observations of Extended Zodiacal
Structures. {\it Ap.J} {\bf 334}, L55-L58.

\item{\sc Sykes, M.V.} 1990. Zodiacal Dust Bands: Their Relation to Asteroid 
Families. {\it Icarus} {\bf 84}, 267-289.

\item{\sc Sykes, M.V. and Greenberg, R.} 1986. The Formation and Origin of the
IRAS Zodiacal Dust Bands as a Consequence of Single Collisions between
Asteroids. {\it Icarus} {\bf 65}, 51-69. 

\item{\sc Sykes, M.V. and Walker, R.G.} 1992. Cometary dust trails. I - Survey.
{\it Icarus} {\bf 95}, 180-210. 

\item{\sc Trulsen, J. and Wikan, A.} 1980. Numerical Simulation of
Poynting-Robertson and Collisional Effects in the Interplanetary Dust Cloud.
{\it Astron. Astrophys.} {\bf 91}, 155-160.

\item{\sc Wallis, M.K. and Hassan, M.H.A.} 1985. Stochastic diffusion of 
interplanetary dust grains orbiting under Poynting-Robertson forces. 
{\it Astron. Astrophys.} {\bf 151}, 435-441.

\item{\sc Whipple, F.L.} 1967. On maintaining the meteoritic complex.
{\it Smithsonian Astrophysical Observatory Special Report} {\bf 239}, 1-46.

\item{\sc Zappala, V., Bendjoya, P.H., Cellino, A., Farinella, P. and 
Froeschle, C.} 1995. Asteroid families: Search of a 12,487 asteroid sample 
using two different clustering techniques. {\it Icarus} {\bf 116}, 291-314.

\end{list}

\clearpage

Table 1

\vspace{20pt}

\begin{tabular}{ccccc}
Asteroid family & $a, \Delta a$ & $e, \Delta e$ & $i, \Delta i$ & 
Area $(10^{9} km^{2})$ \\
Eos & 3.015, 0.012 & 0.076, 0.009 & 9.35, 1.5 & 4.0 \\
Themis & 3.148, 0.035 & 0.155, 0.013 & 1.43, 0.32 & 0.35 \\
Koronis & 2.876, 0.026 & 0.047, 0.006 & 2.11, 0.09 & 0.35 \\
\end{tabular}

\clearpage

\begin{figure}[t]
 \centering
 \centerline{
 \psfig{file=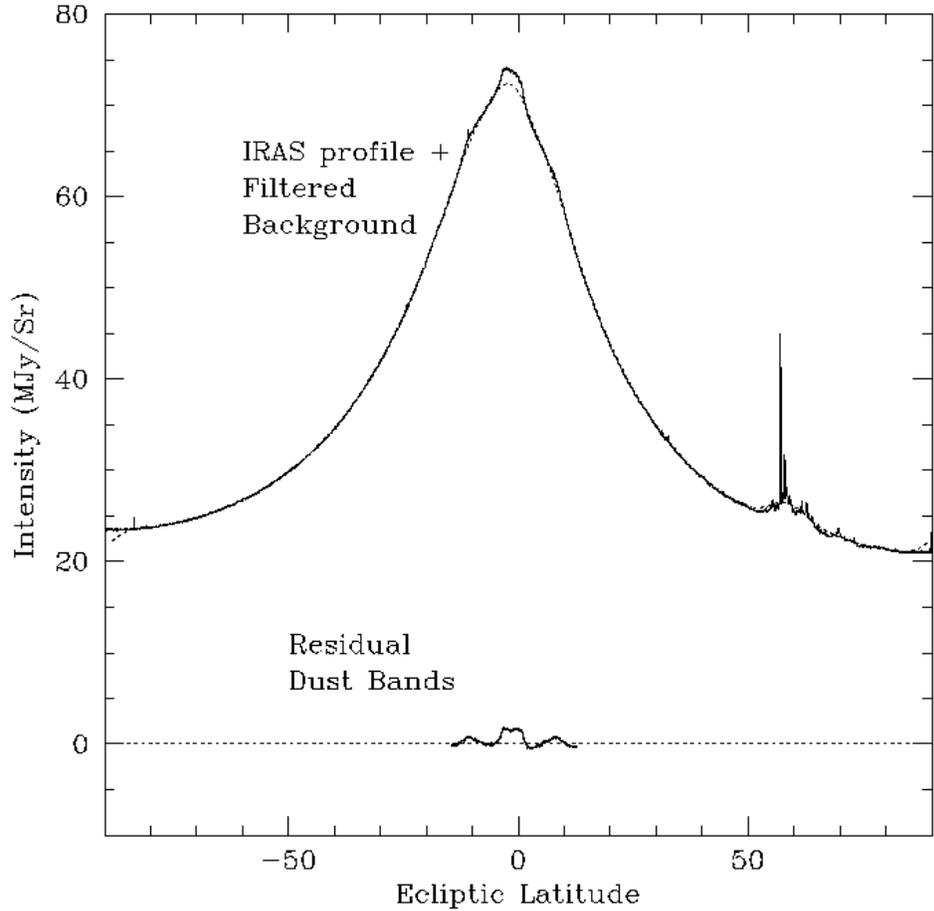,width=5in}
            }
 \caption{IRAS observation of the zodiacal cloud in the 25 $\mu m$ waveband 
(solid curve). This observation was made at 90$^{\circ}$ solar elongation 
angle in the direction leading the Earth in its orbit when the planet was at 
an ecliptic longitude of 293$^{\circ}$. The dust bands can be seen as 
projecting "shoulders" near $\pm$10$^{\circ}$ and 0$^{\circ}$ latitude. The 
structure around 60$^{\circ}$ latitude is due to dust in the plane of the 
Galaxy. By applying a Fourier filter to the IRAS observation a smooth 
background profile (underlying dotted curve) is separated from a high 
frequency dust band profile (solid lower curve). This filtered high frequency 
dust band profile is merely a residual representing the "tip of the iceberg" 
in terms of dust band material in the zodiacal cloud.}
\end{figure}

\clearpage

\begin{figure}[t]
 \centering
 \centerline{
 \psfig{file=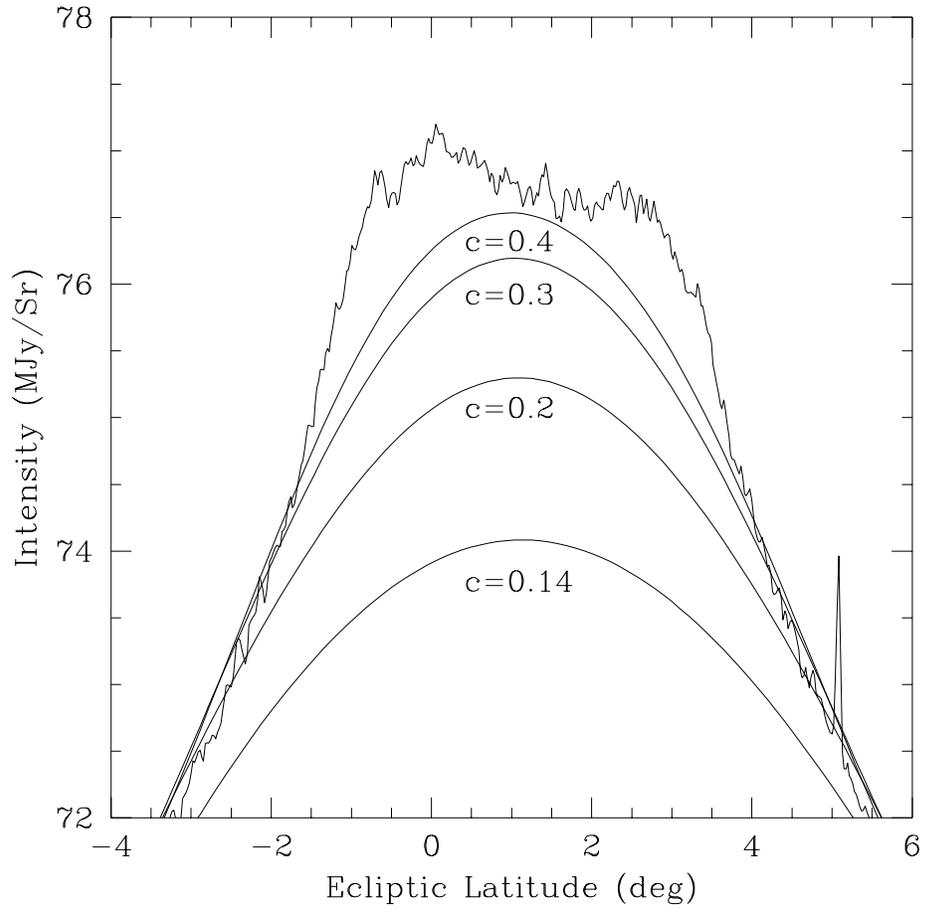,width=5in}
            }
 \caption{The coefficient $c$ represents the fraction of frequency points to 
remain after the high frequencies are stripped from the Fourier transform. 
This figure shows how more and more of the original profile is incorporated 
into the low frequency background as the constant $c$ increases. This is a 
dramatic illustration of the arbitrary nature of any filtering process and the 
danger of assuming the resultant residuals to represent the complete dust band 
structure.}
\end{figure}

\clearpage

\begin{figure}[t]
 \centering
 \centerline{
 \psfig{file=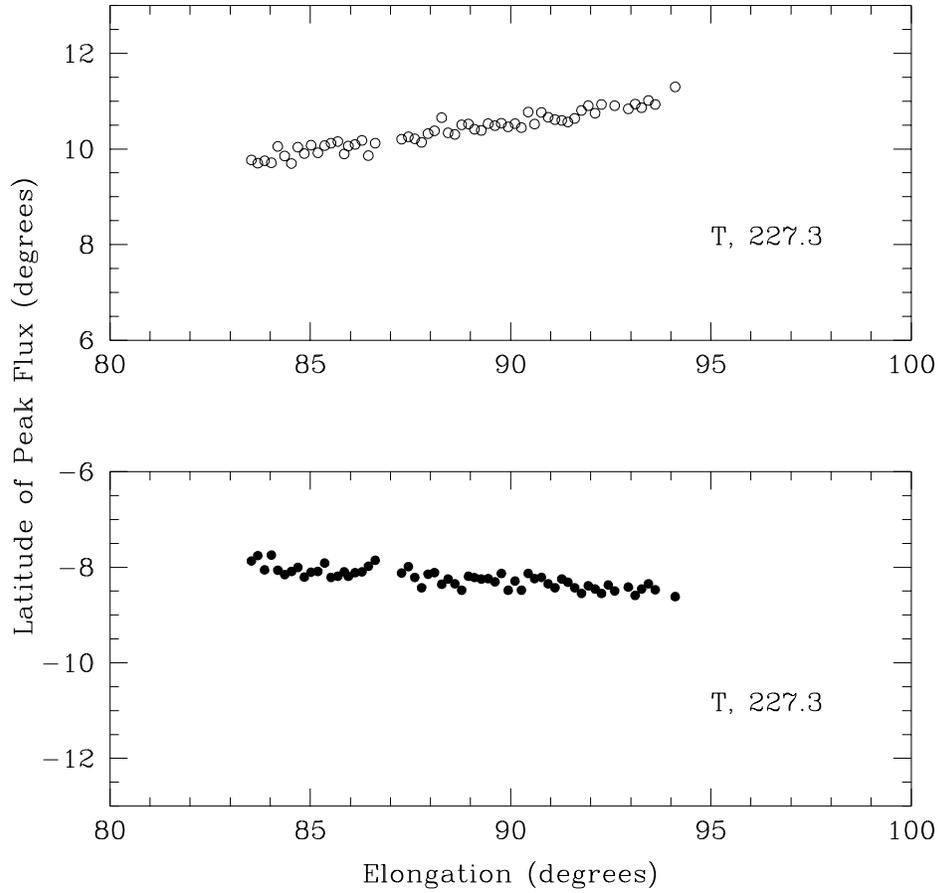,width=5in}
            }
 \caption{The variation with solar elongation angle of the latitude of peak 
flux of the north and south ten degree bands, at a longitude of Earth of 
227.3$^{\circ}$ in the trailing direction. As the solar elongation angle
increases, the effective distance to the dust band material decreases, and
parallax causes the observed peak latitudes to increase. Over a small range
of solar elongation angle such as that shown, this effect can be assumed to
be linear.}
\end{figure}

\clearpage

\begin{figure}[t]
 \centering
 \centerline{
 \psfig{file=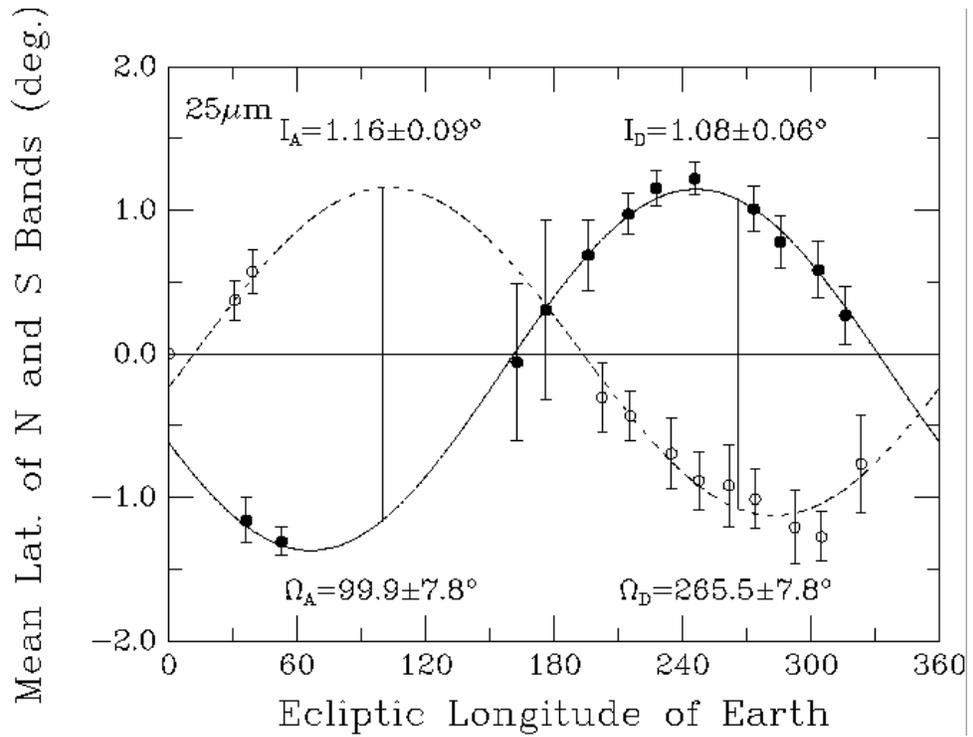,width=5in}
            }
 \caption{The variation of the mean of the north and south ten degree band peak 
latitudes in the 25 $\mu m$ waveband with ecliptic longitude of Earth. The 
sinusoidal variation indicates that the plane of symmetry of the bands is 
inclined to the ecliptic. The amplitude of the variation is slightly more than 
one degree - close to the inclination of Jupiter's orbit. This is an indication 
that the material associated with the bands is located in the asteroid belt.}
\end{figure}

\clearpage
\newpage
\pagebreak

\begin{figure}
 \begin{center}
 \begin{tabular}{l}
 \psfig{file=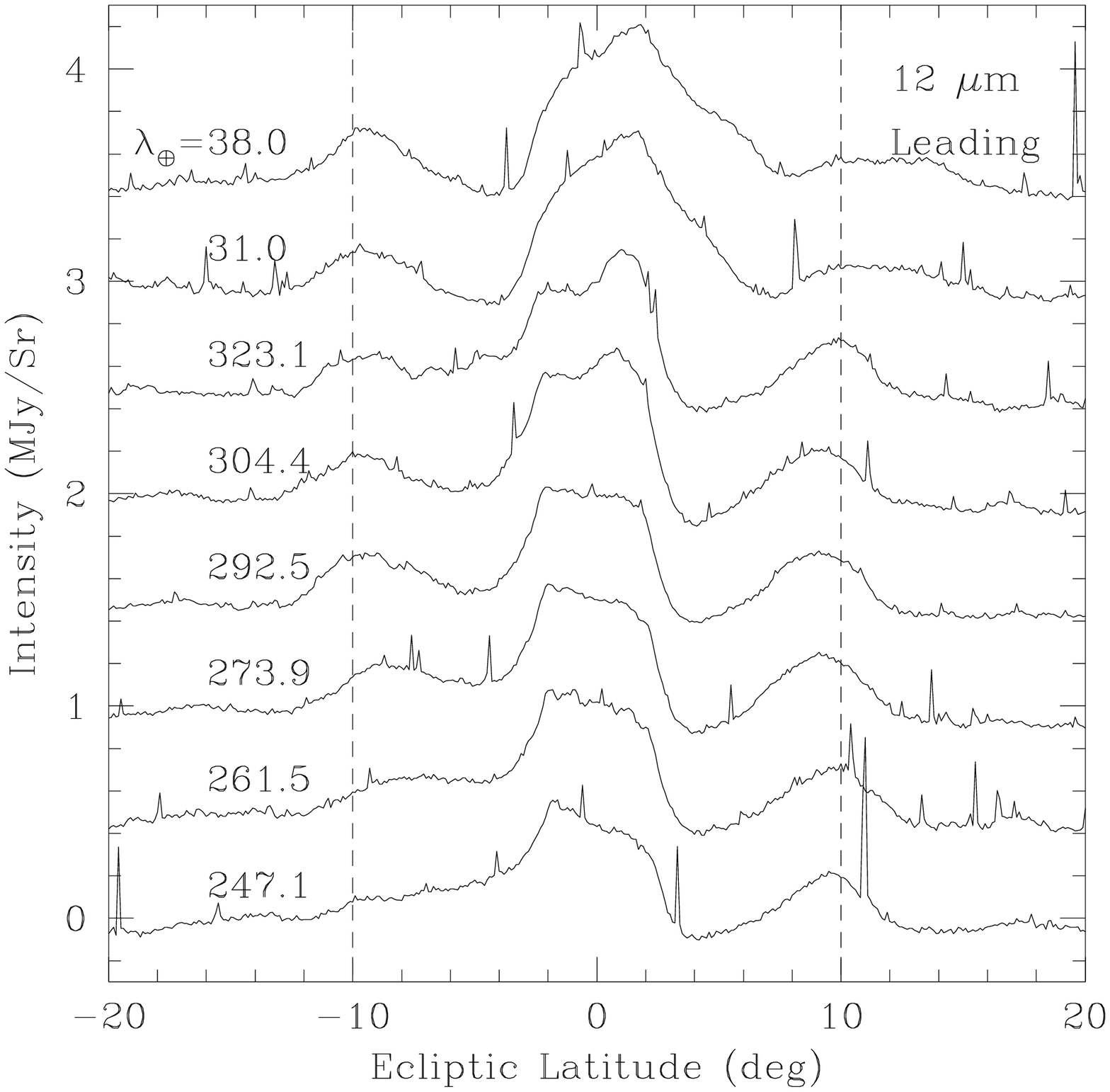,width=3in} \\
 \psfig{file=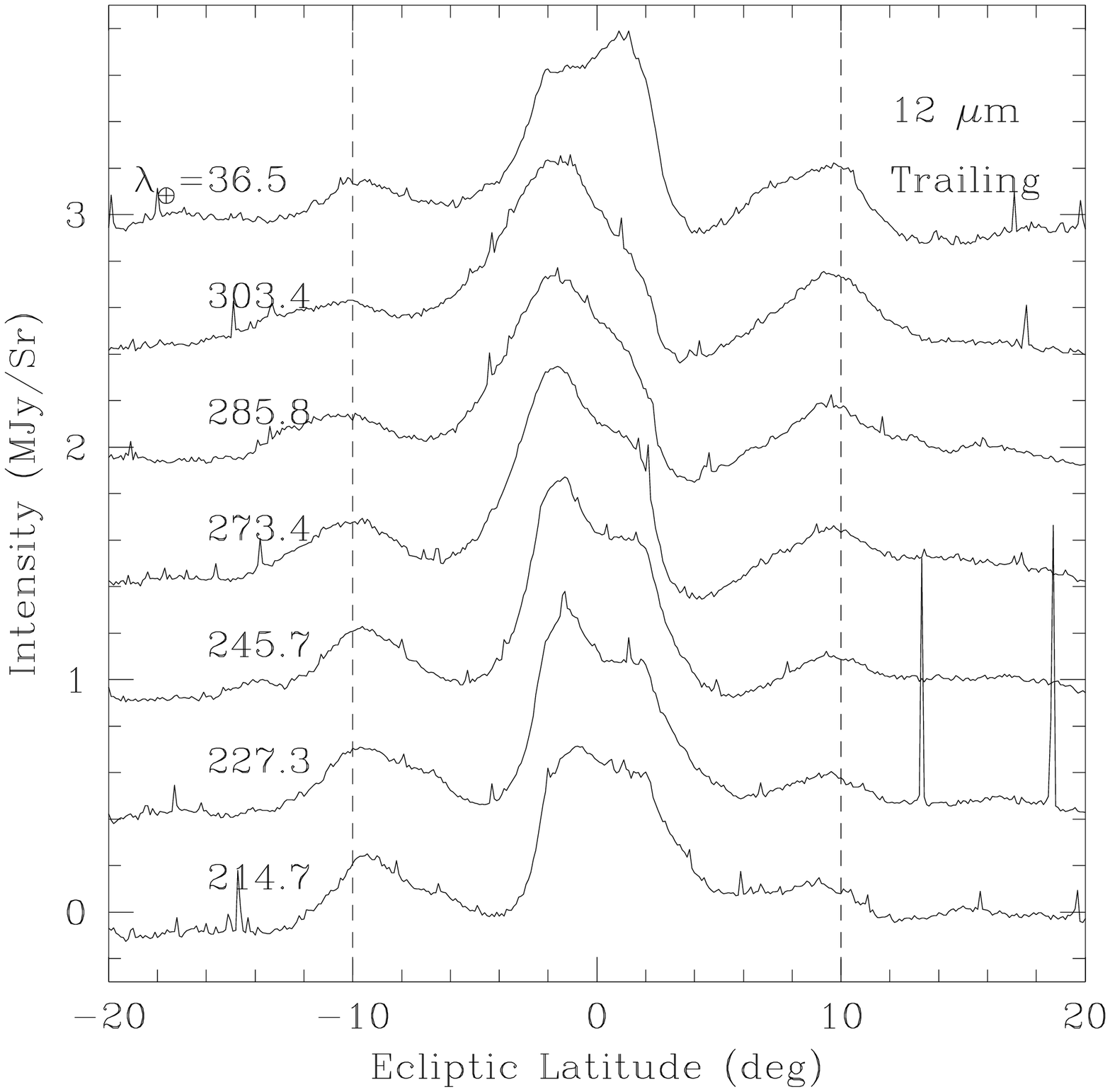,width=3in} \\
 \end{tabular}
 \end{center}
 \caption{Coadded IRAS dust band profiles as seen in the 12 $\mu m$ waveband 
normalized to a solar elongation of 90$^{\circ}$ in the leading (top) and 
trailing (bottom) directions. Dashed vertical lines are placed at 
$\pm$10$^{\circ}$ for reference.}
\end{figure}

\clearpage

\begin{figure}
 \begin{center}
 \begin{tabular}{l}
 \psfig{file=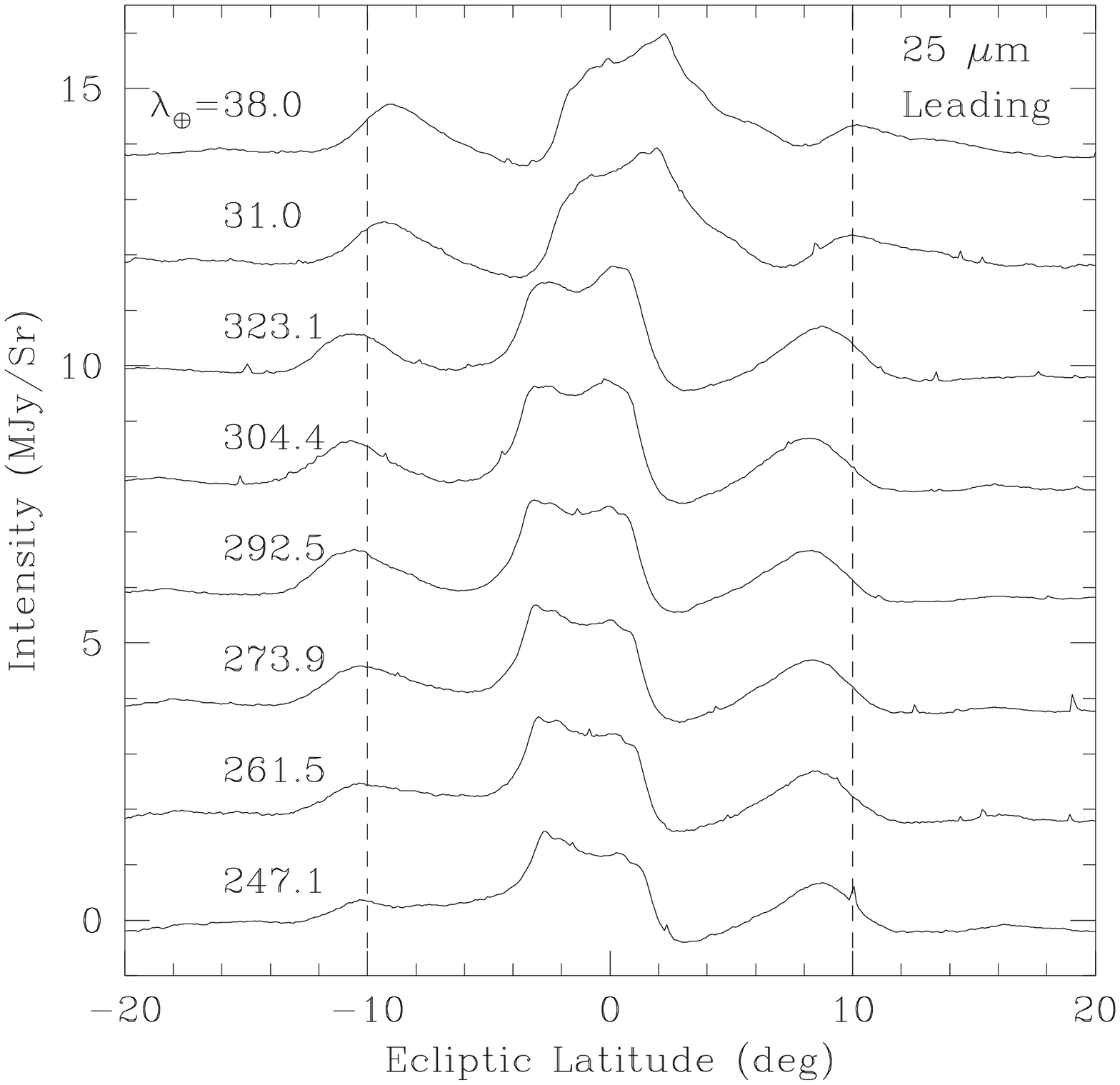,width=3in} \\
 \psfig{file=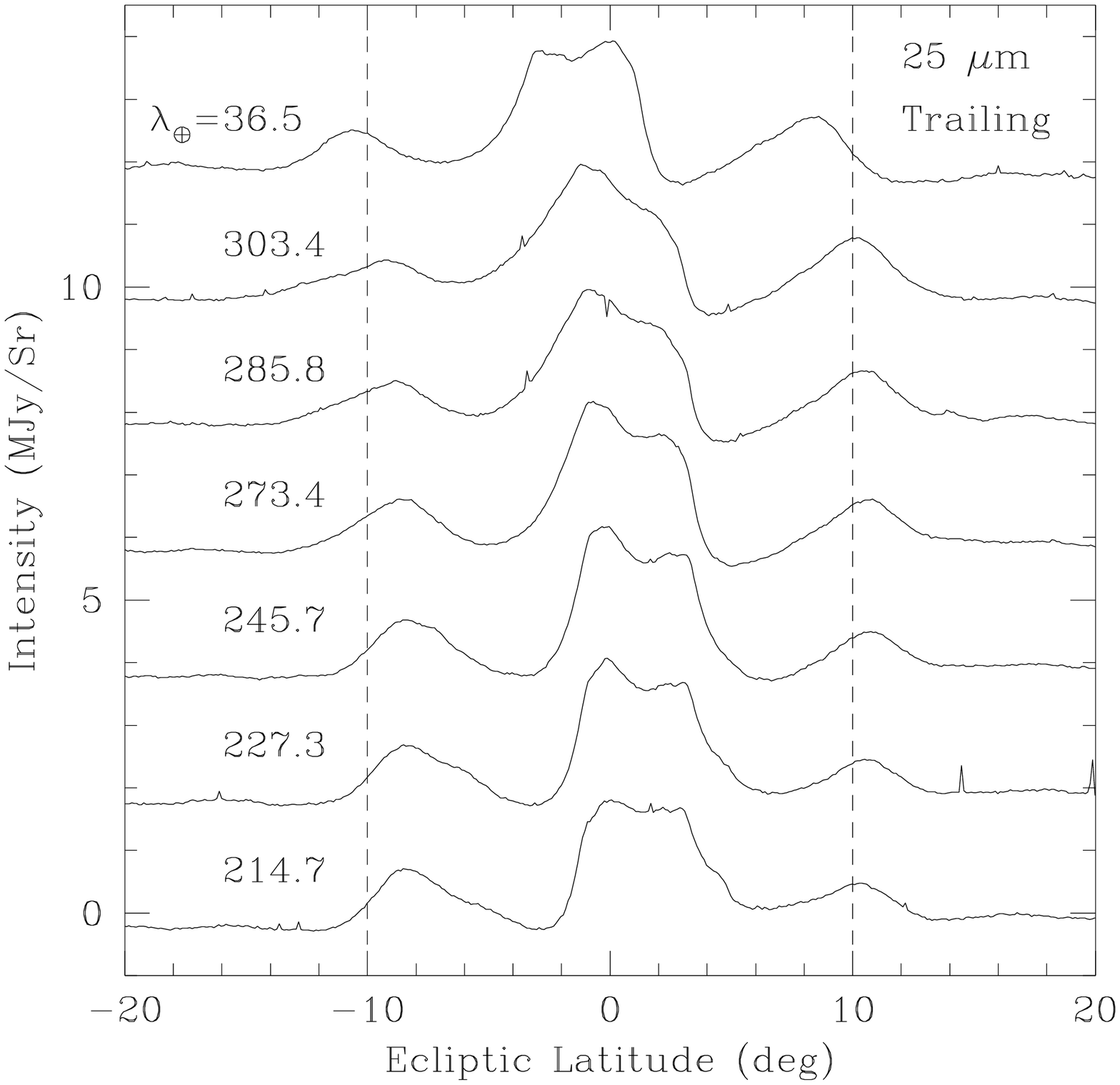,width=3in} \\
 \end{tabular}
 \end{center}
 \caption{Coadded IRAS dust band profiles as seen in the 25 $\mu m$ waveband 
normalized to a solar elongation of 90$^{\circ}$ in the leading (top) and 
trailing (bottom) directions. Dashed vertical lines are placed at 
$\pm$10$^{\circ}$ for reference.}
\end{figure}

\clearpage

\begin{figure}
 \begin{center}
 \begin{tabular}{l}
 \psfig{file=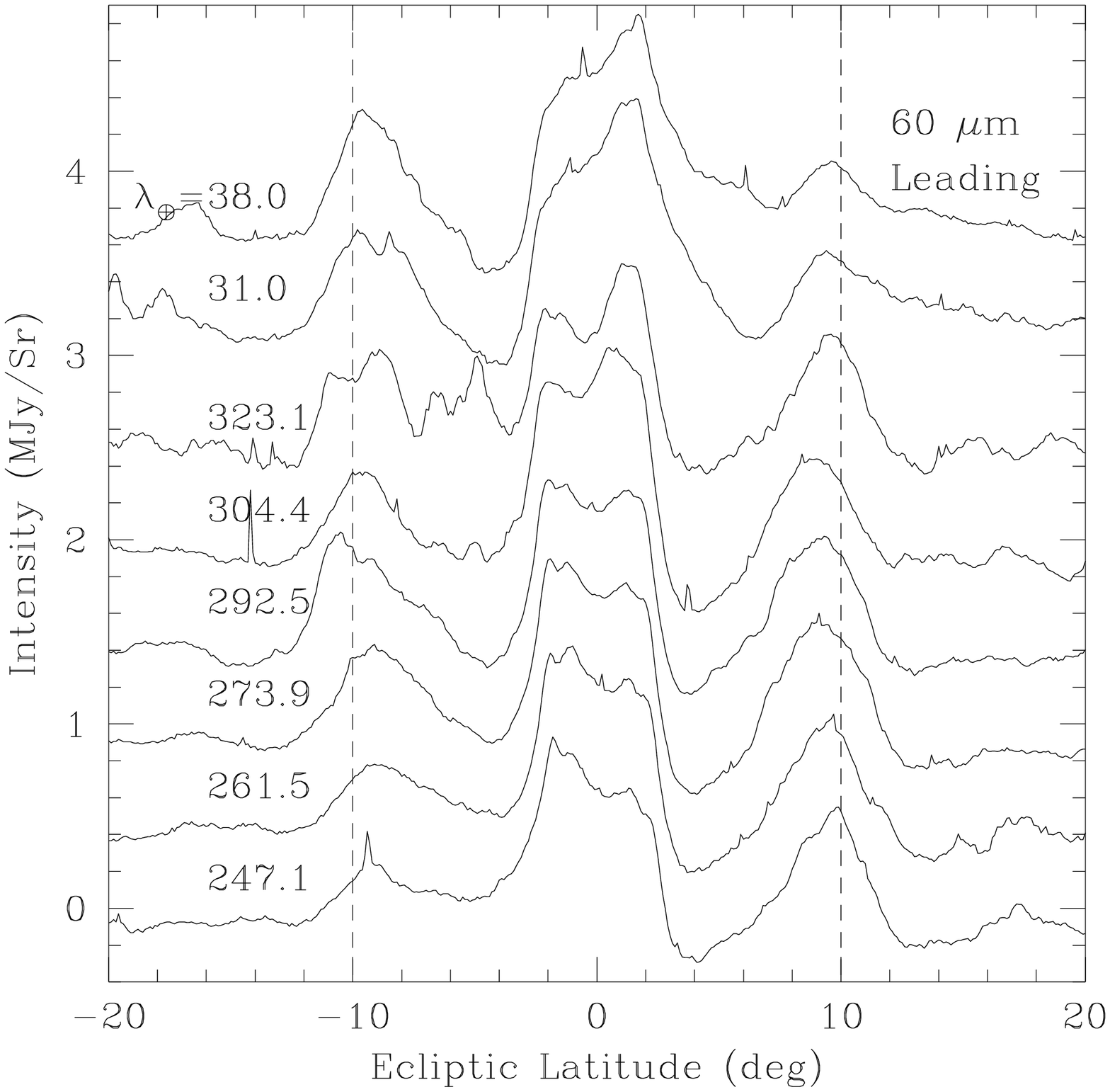,width=3in} \\
 \psfig{file=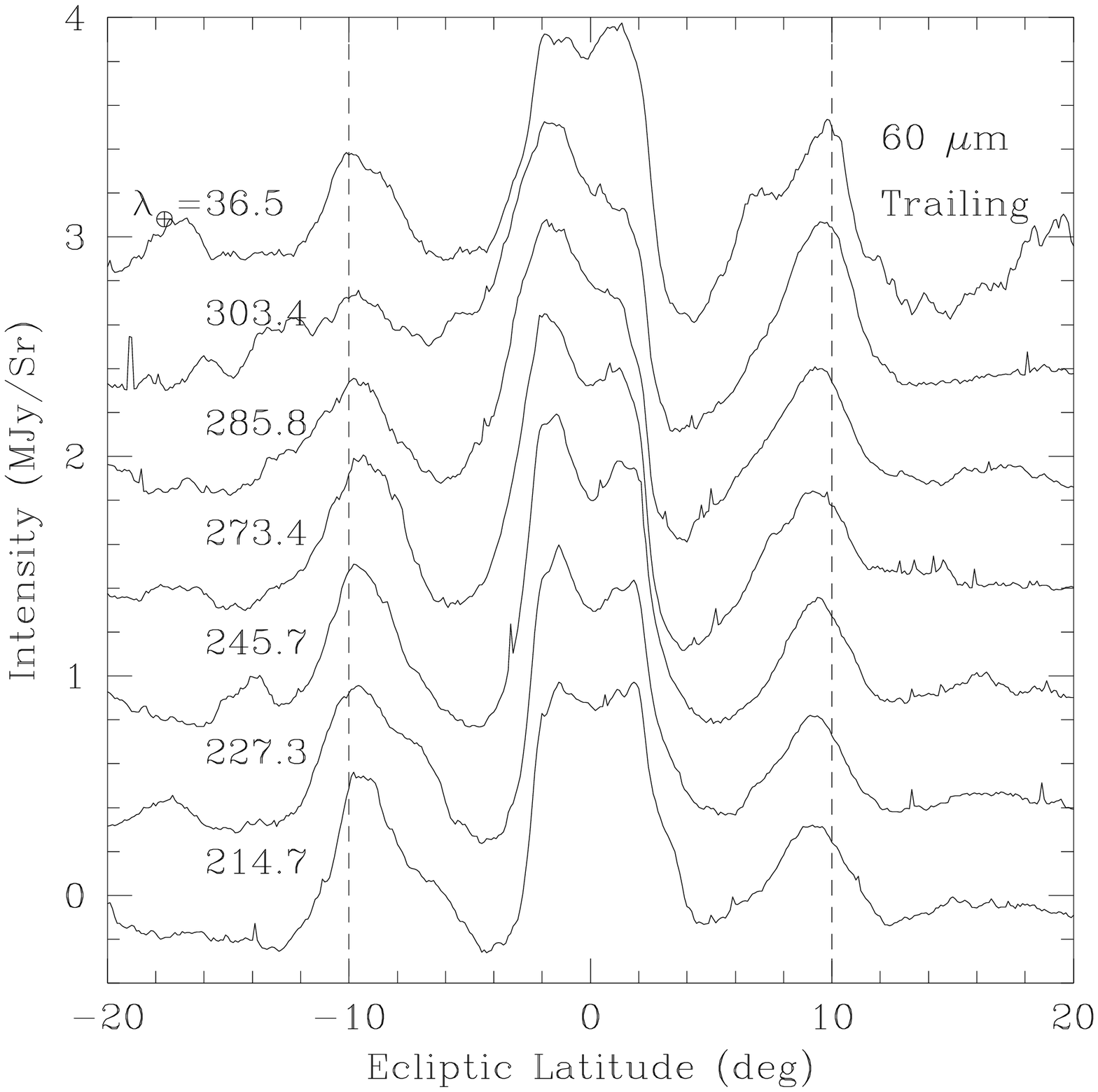,width=3in} \\
 \end{tabular}
 \end{center}
 \caption{Coadded IRAS dust band profiles as seen in the 60 $\mu m$ waveband 
normalized to a solar elongation of 90$^{\circ}$ in the leading (top) and 
trailing (bottom) directions. Dashed vertical lines are placed at 
$\pm$10$^{\circ}$ for reference.}
\end{figure}

\clearpage

\begin{figure}[t]
 \centering
 \centerline{
 \psfig{file=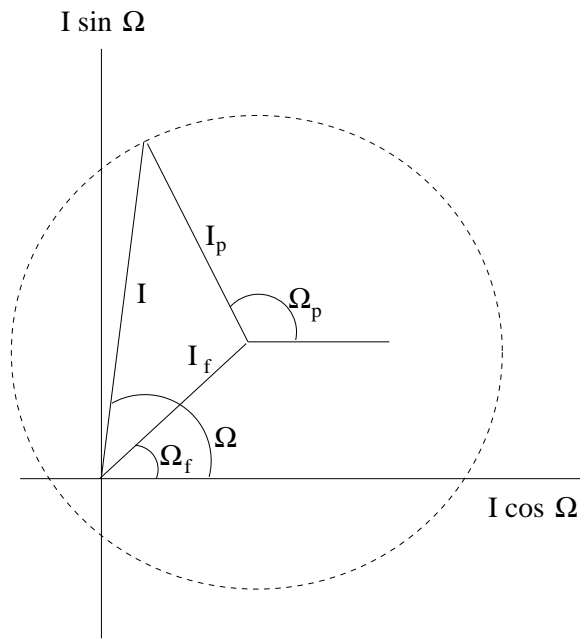,width=3in}
            }
 \caption{The vectorial relationship between the osculating, proper and forced 
elements in  ($I$ cos $\Omega$, $I$ sin $\Omega$) space. The distributions are 
displaced from the origin due to 
the forced elements and the radius of the distributions represents the proper 
elements. An equivalent relationship exists for eccentricity and pericenter.}
\end{figure}

\begin{figure}[t]
 \centering
 \centerline{
 \psfig{file=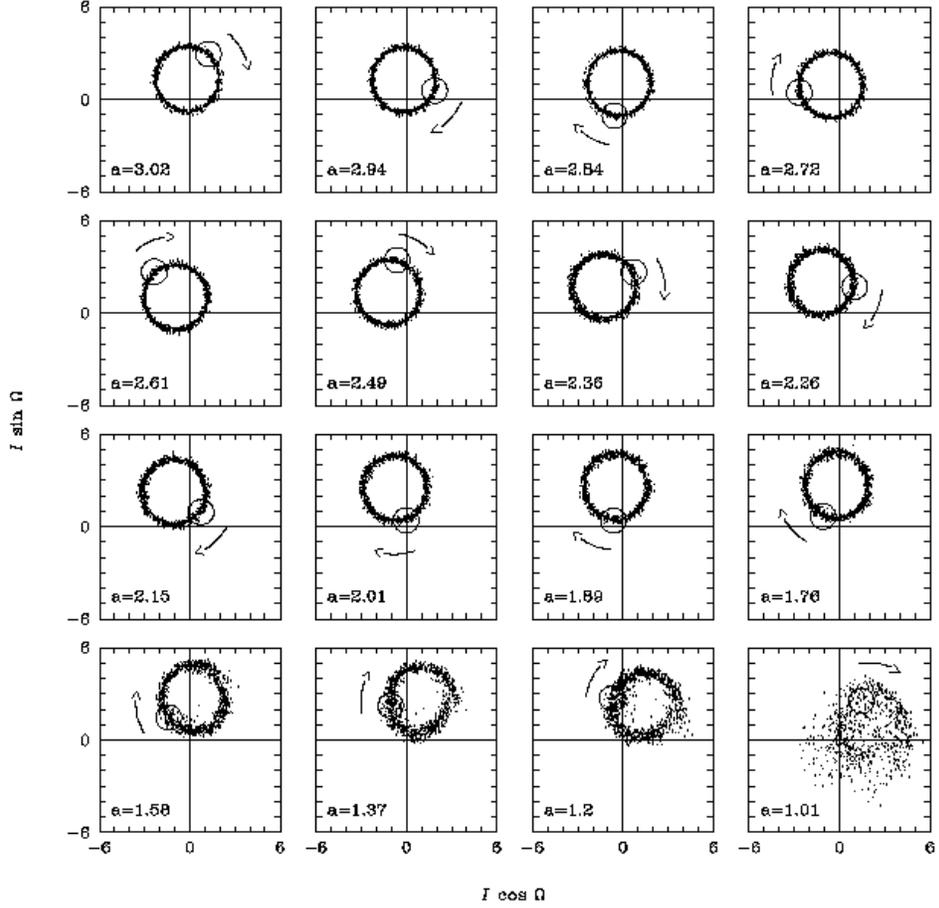,width=5in}
            }
 \caption{Sixteen frames showing the inclination distribution of a wave of 
 10 $\mu m$ diameter Koronis dust particles migrating from the asteroid belt
 toward the Sun (Kortenkamp and Dermott 1998). The particles were released 
 52000 years ago with a mean semi-major axis of 3.02 AU. The wave reached
 1.01 AU at the present time, indicating approximately 3500 years between
 each frame. The motion of the `center of mass' of the distribution 
 represents the changing $I_{f}$ and $\Omega_{f}$ of the wave. The proper
 inclination $I_{p}$ manifested as the radius of the circular distribution
 remains constant until the wave approaches Earth and many particles are
 gravitationally scattered. The large open circle in each frame follows the
 evolution of a single orbit and reveals the clockwise precession of 
 $\Omega_{p}$.}
\end{figure}

\begin{figure}[t]
 \centering
 \centerline{
 \psfig{file=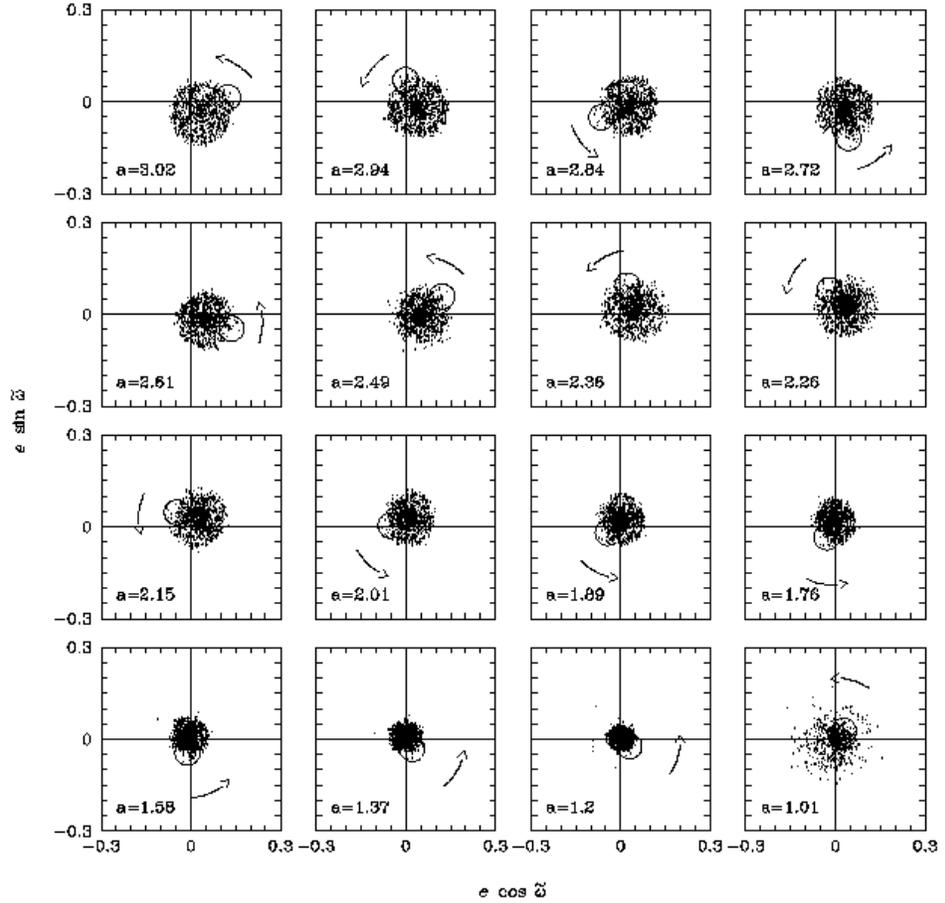,width=5in}
            }
 \caption{Same as the previous figure, but now the eccentricity distribution
 of the migrating wave of Koronis particles is shown. After the wave passes
 through the $\nu_{6}$ secular resonance near 2.2 AU the decay in the proper
 and forced eccentricity becomes more evident. Gravitational scattering by
 the Earth and trapping in exterior mean motion resonances with the planet
 cause the eccentricities of some of the orbits to increase as they approach
 1 AU. The large open circle in each frame follows the evolution of one 
 particular orbit and reveals the counter-clockwise precession of $\varpi_{p}$.}
\end{figure}

\clearpage

\begin{figure}[t]
 \centering
 \centerline{
 \psfig{file=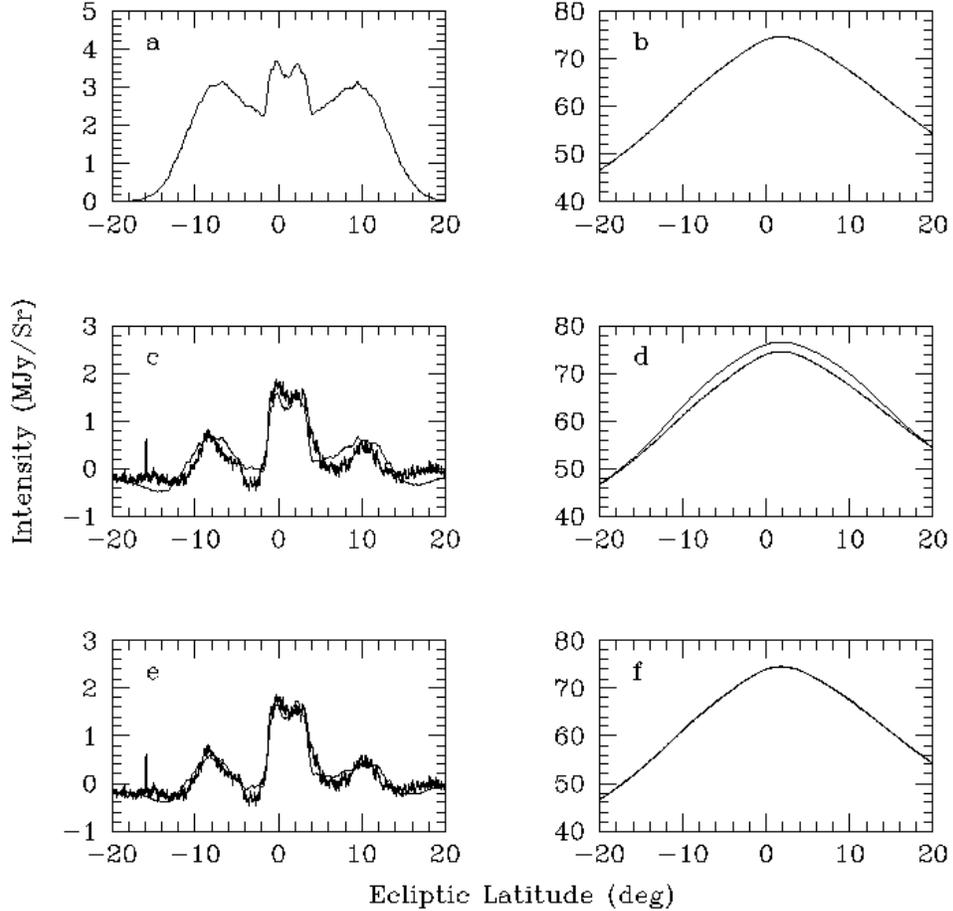,width=5in}
            }
 \caption{The iterative filtering procedure. Panel (a) shows a raw model dust 
band having the same viewing geometry as an observed background (b). In the 
first iteration (a) is added to (b) and the sum is filtered to obtain (c), a 
model dust band (smooth curve); the observed dust bands (noisy curve) are also 
shown for comparison. The background obtained from this iteration shown in 
panel (d) is of a higher intensity than the original background due to the fact 
that it contains two low frequency dust band components, one from the addition 
of the model dust band and one from the actual dust band in the original 
observed background. In the final iteration we subtract the excess intensity 
shown in (d) from the original background (b) and add (a) before filtering to 
obtain the final dust band (e) and the final background (f) that agree with 
the observations.}
\end{figure}

\clearpage

\begin{figure}
 \begin{center}
 \begin{tabular}{l}
 \psfig{file=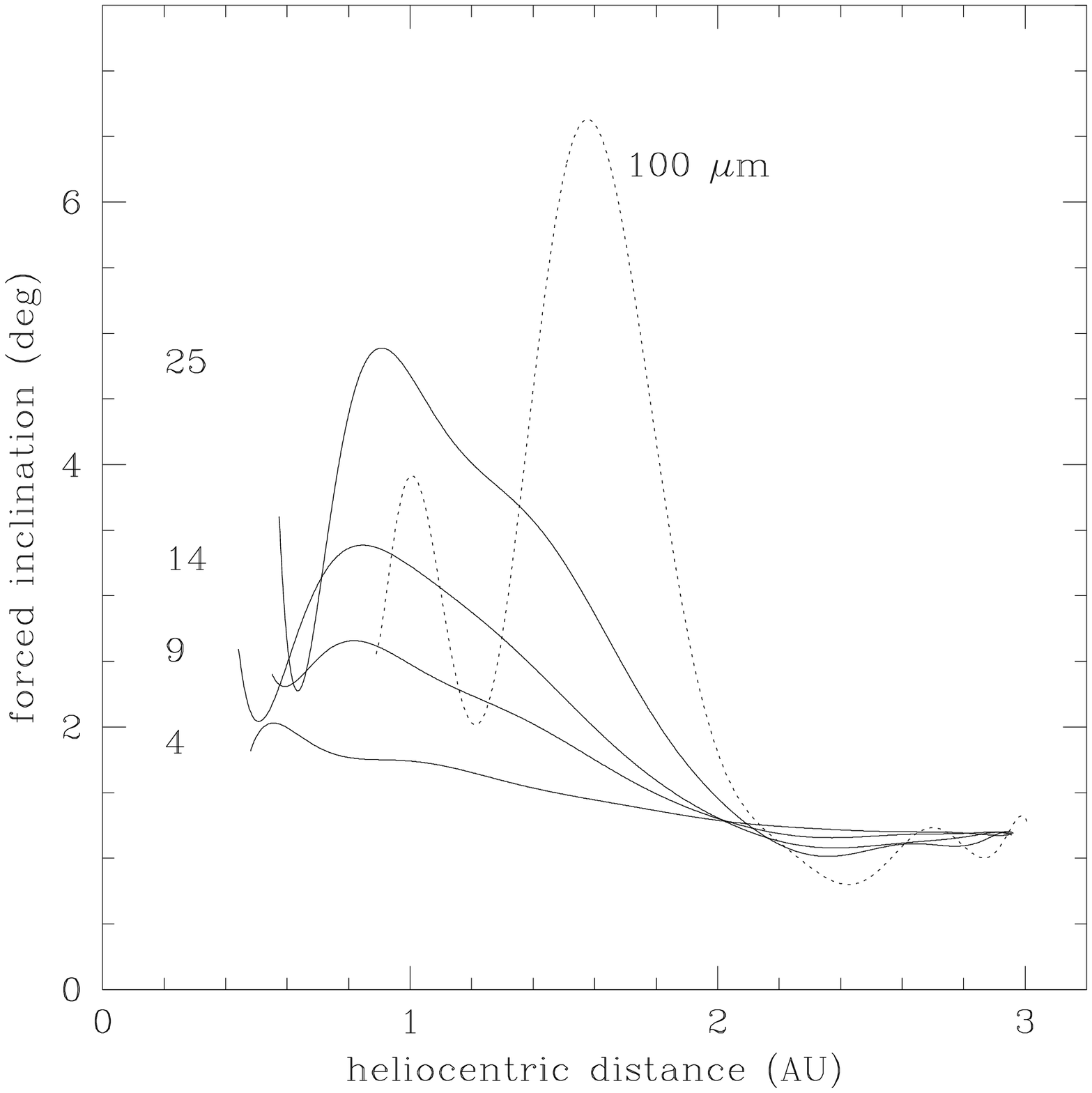,width=3in} \\
 \psfig{file=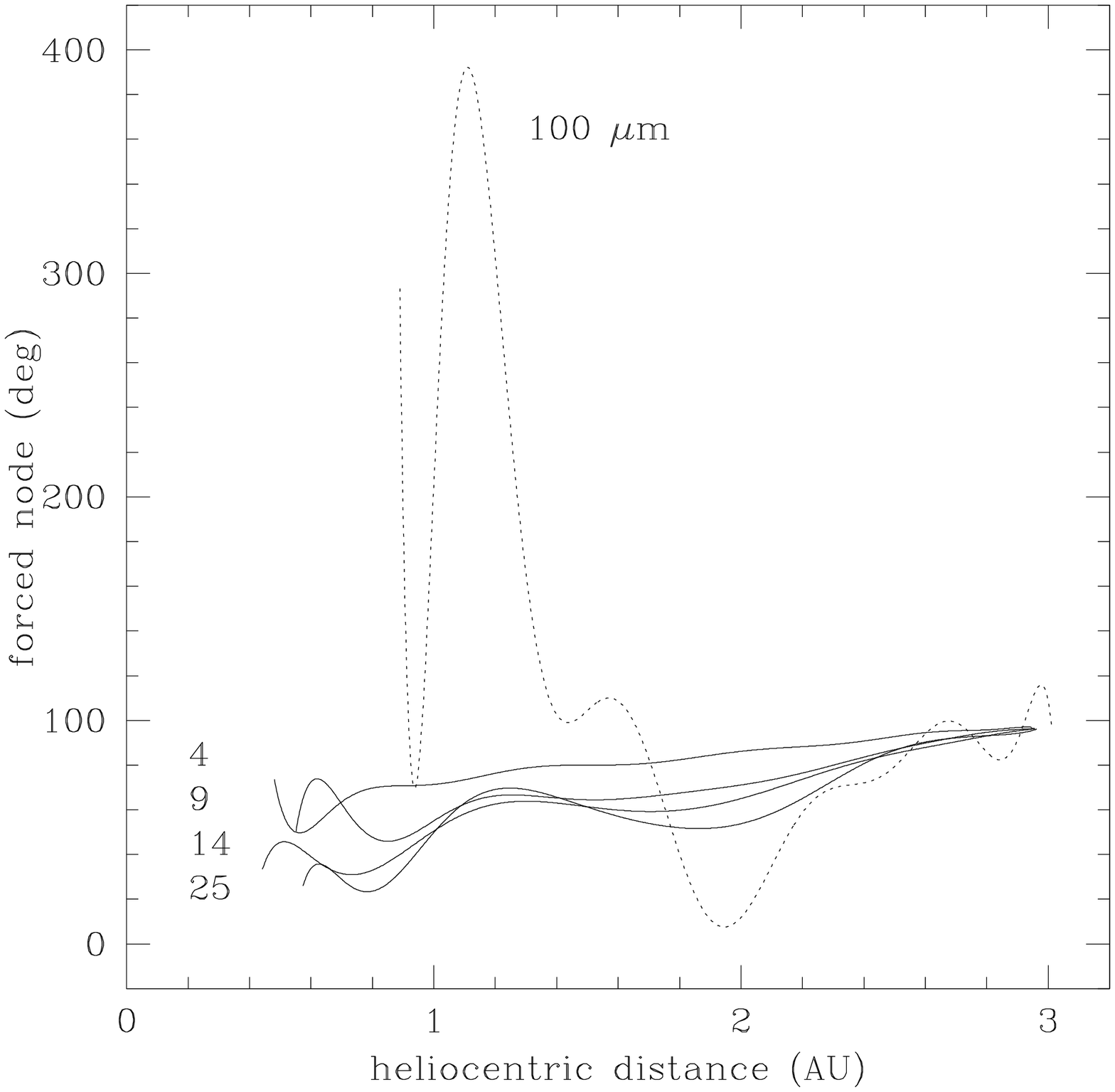,width=3in} \\
 \end{tabular}
 \end{center}          
 \caption{The variation with heliocentric distance of the forced inclination 
(top) and node (bottom) of dust particles in the zodiacal cloud. This variation 
is shown for five different particle sizes. The secular 
resonance near 2 AU has a dramatic effect which increases with particle size: 
larger particles, having longer P-R drag lifetimes, spend a longer amount of 
time in the resonance.}
\end{figure}

\clearpage

\begin{figure}
 \begin{center}
 \begin{tabular}{l}
 \psfig{file=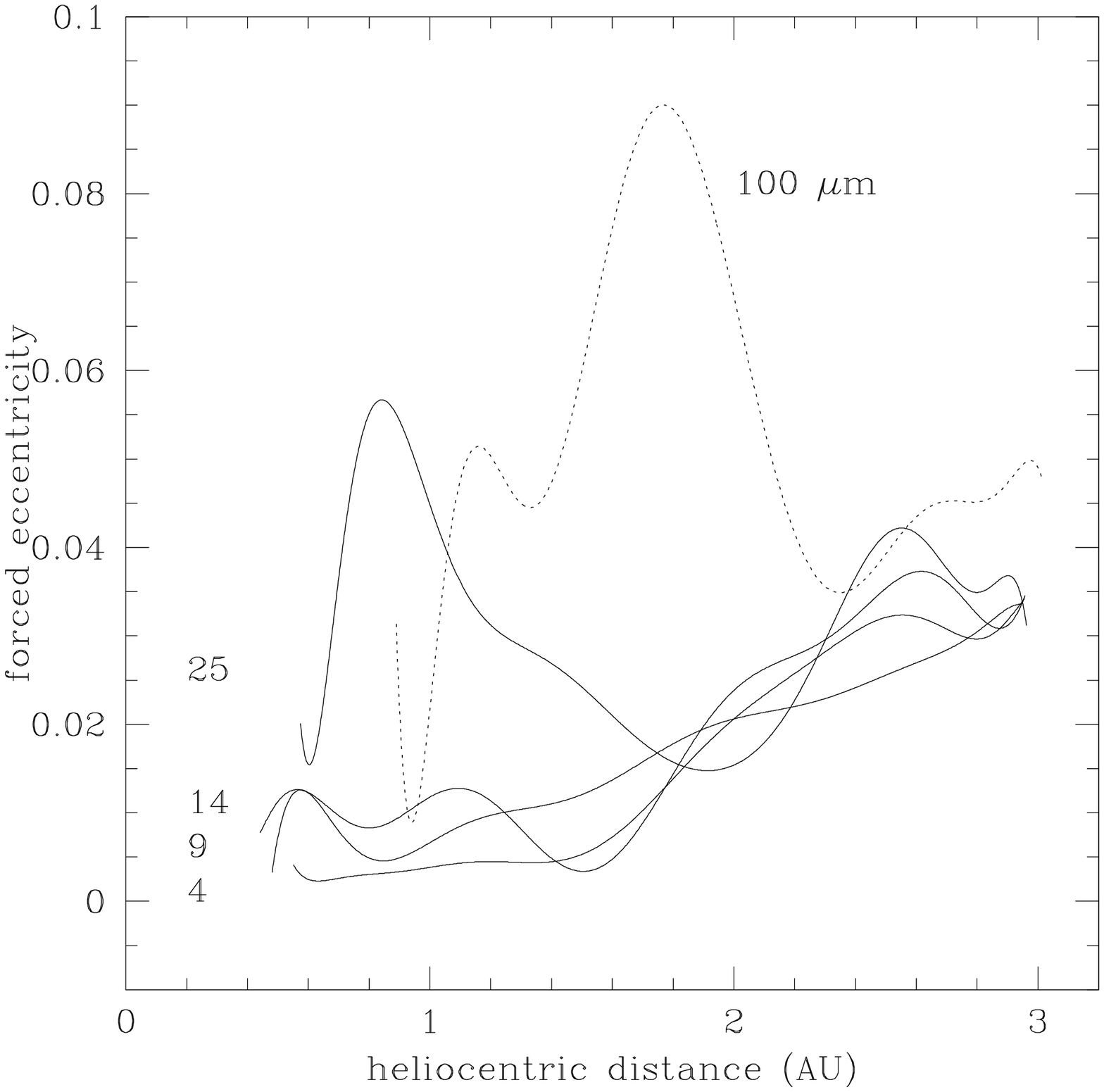,width=3in} \\
 \psfig{file=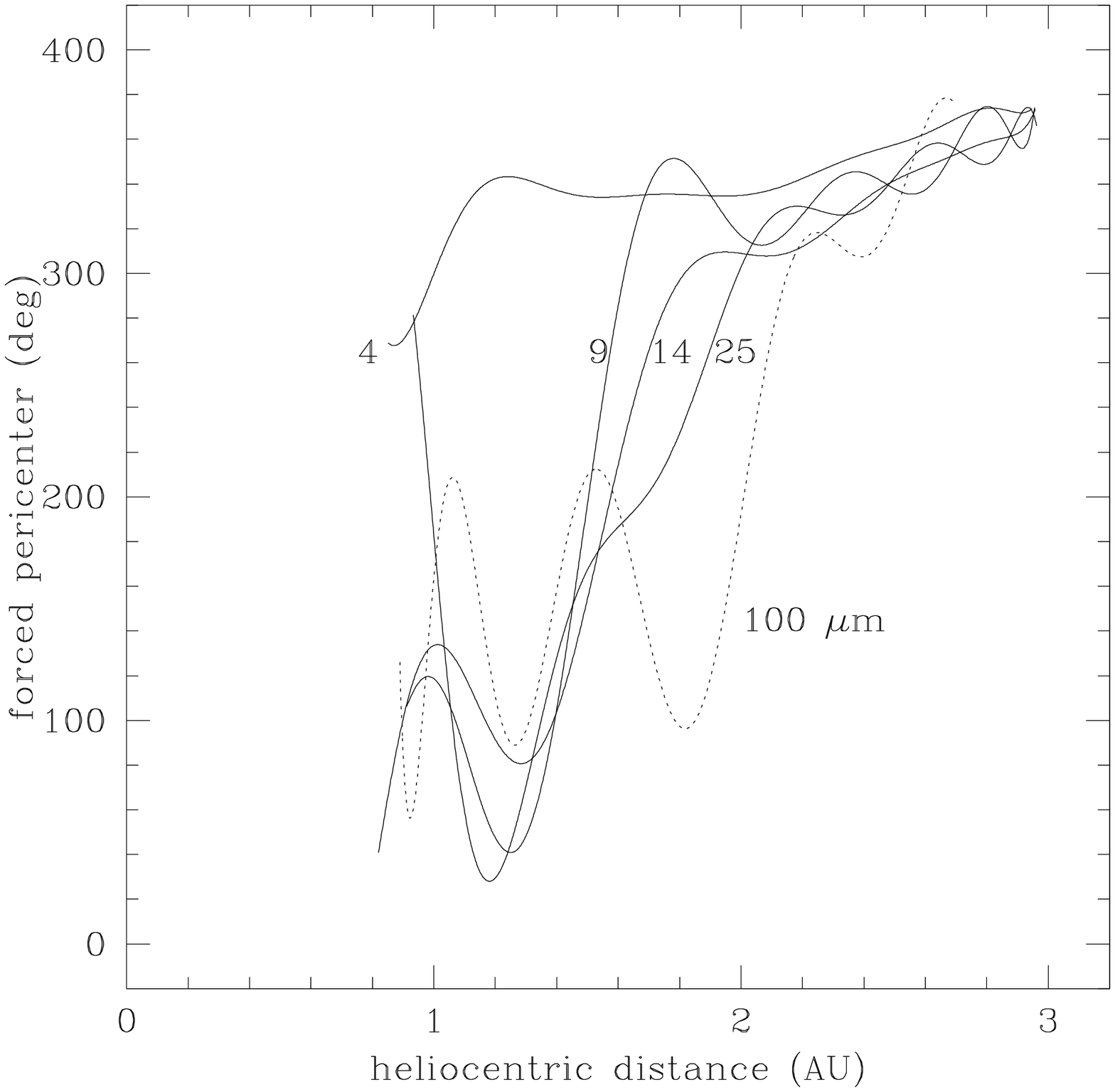,width=3in} \\
 \end{tabular}
 \end{center}
 \caption{The variation with heliocentric distance of the forced eccentricity 
(top) and pericenter (bottom) of dust particles in the zodiacal cloud. This 
variation is shown for five different particle sizes. The secular 
resonance near 2 AU has a dramatic effect which increases with particle size: 
larger particles, having longer P-R drag lifetimes, spend a longer amount of 
time in the resonance.}
\end{figure}

\clearpage

\begin{figure}
 \begin{center}
 \begin{tabular}{l}
 \psfig{file=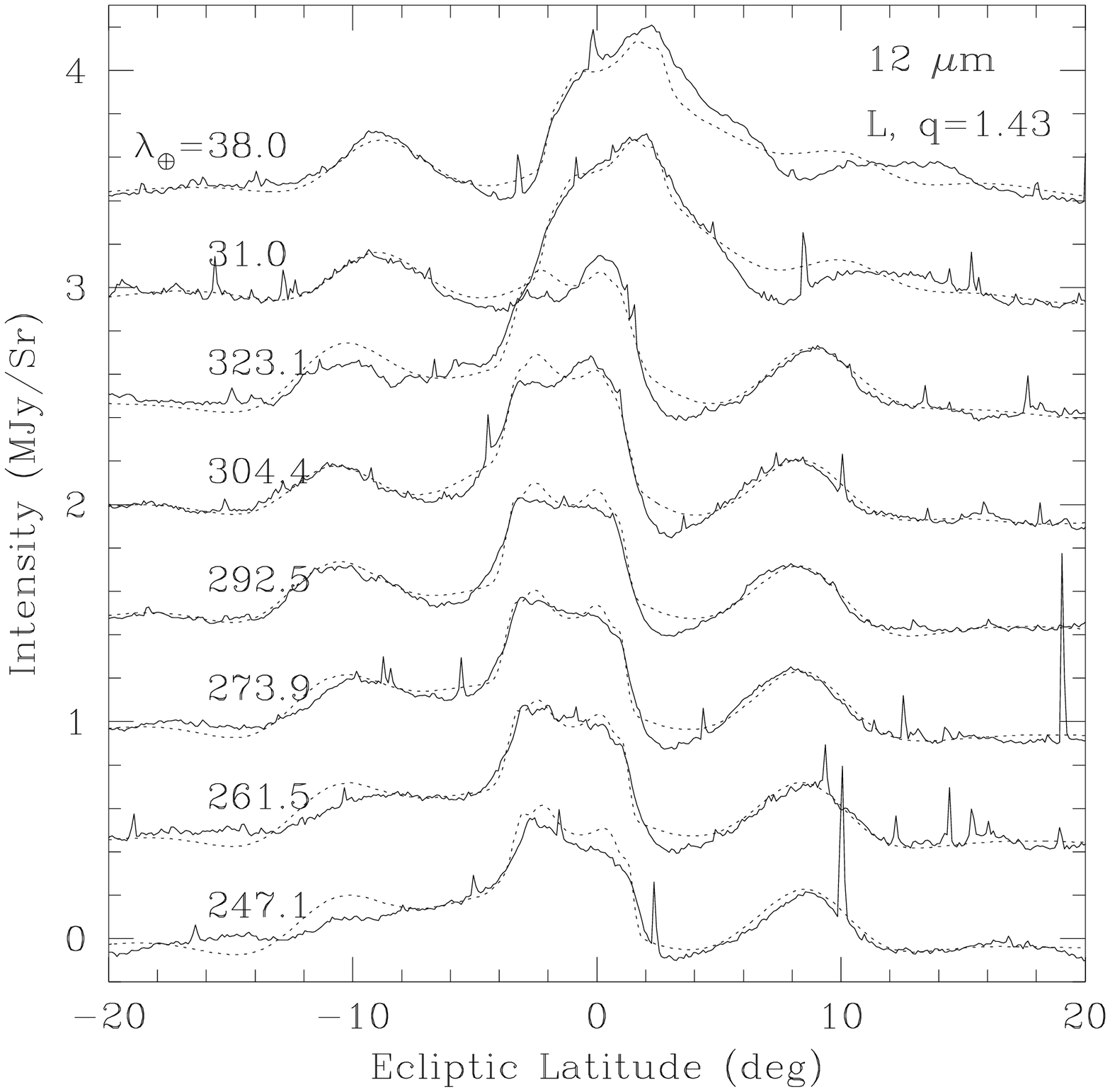,width=3in} \\
 \psfig{file=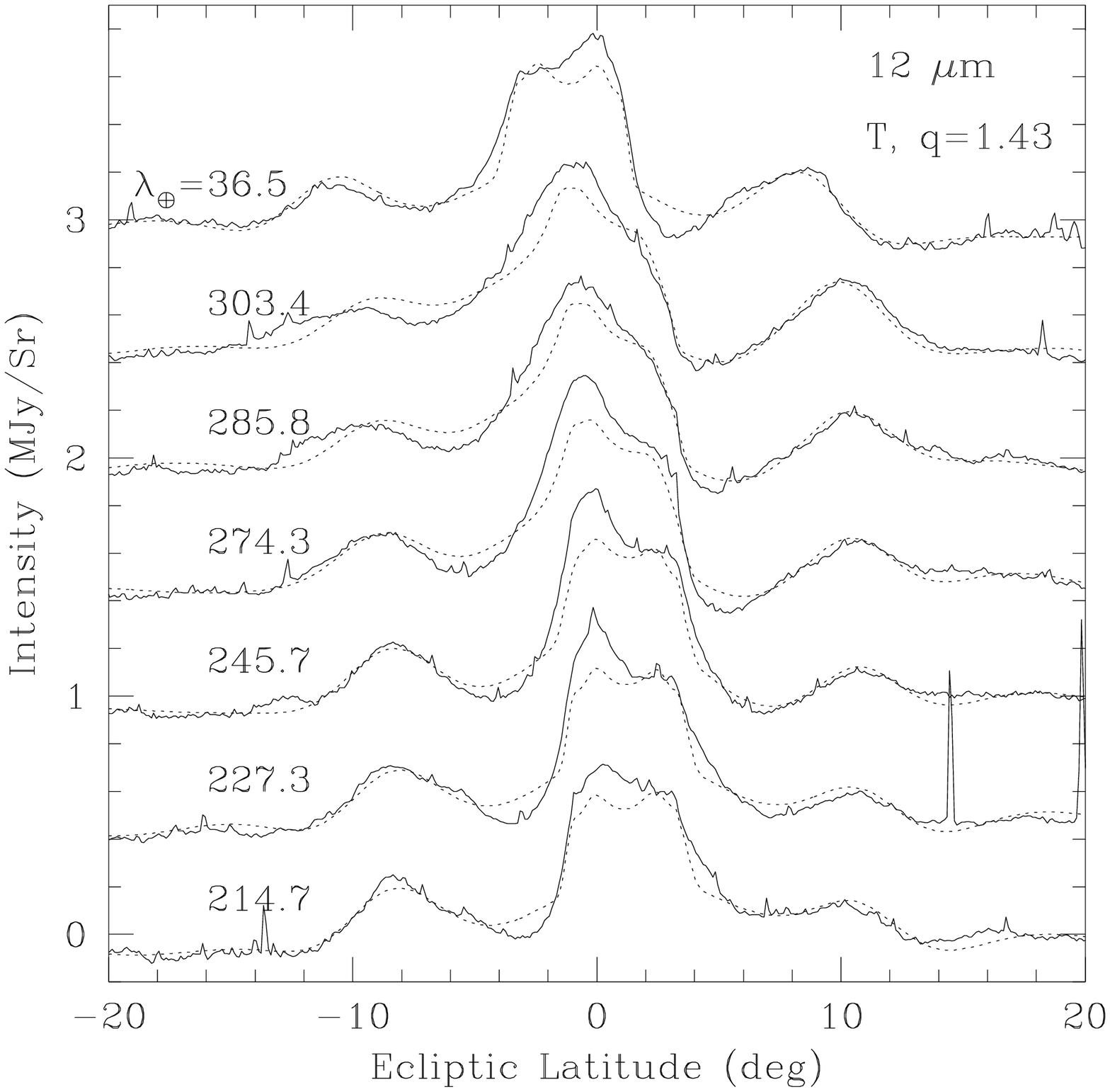,width=3in} \\
 \end{tabular}
 \end{center}
 \caption{Best fit dust band model (dotted curves) to the 12 $\mu m$ IRAS 
observations in both the leading (top) and trailing (bottom) directions. This 
model has a size-frequency index $q$ of 1.43, indicating that large particles 
dominate the distribution.}
\end{figure}

\clearpage

\begin{figure}
 \begin{center}
 \begin{tabular}{l}
 \psfig{file=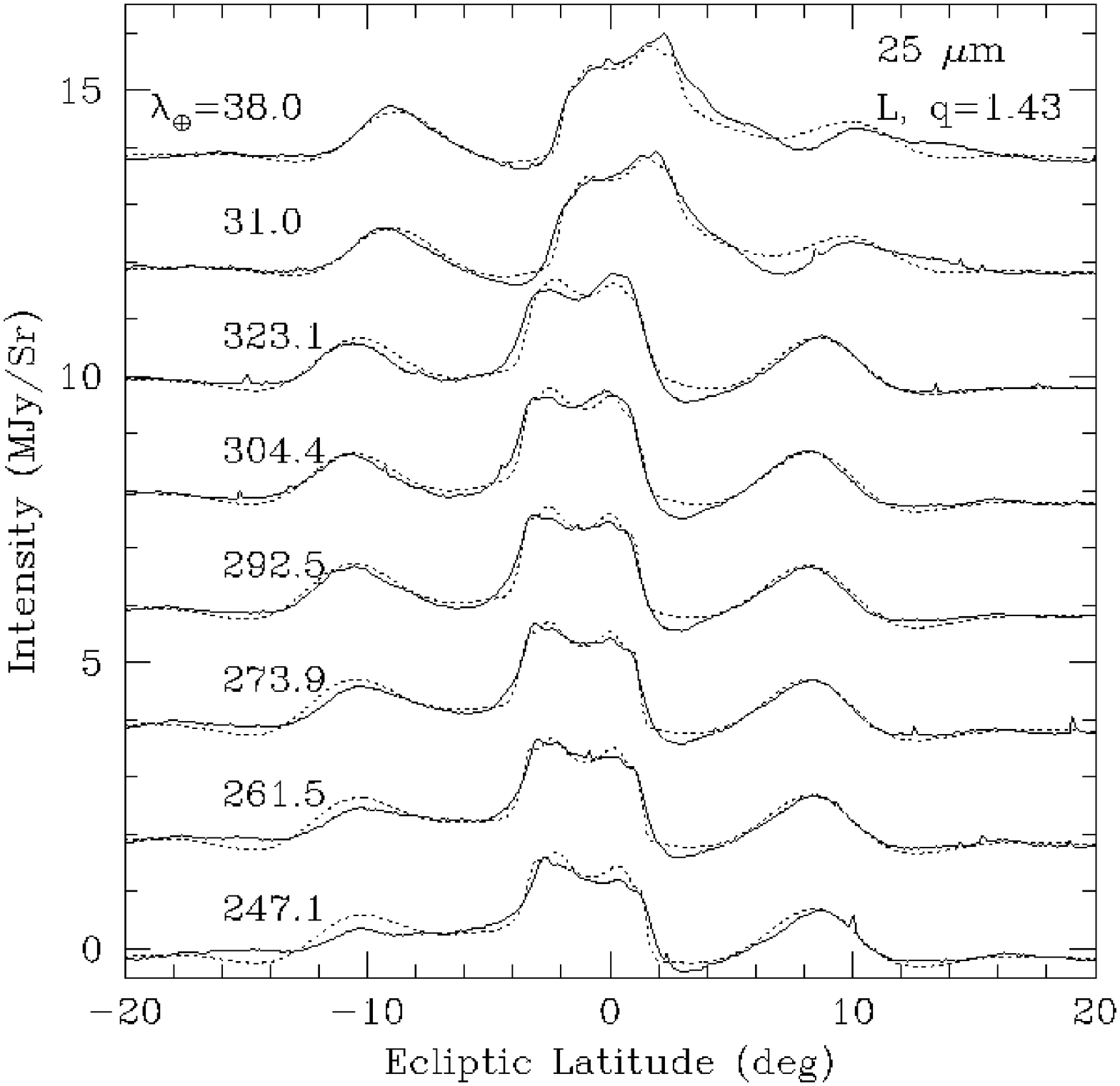,width=3in} \\
 \psfig{file=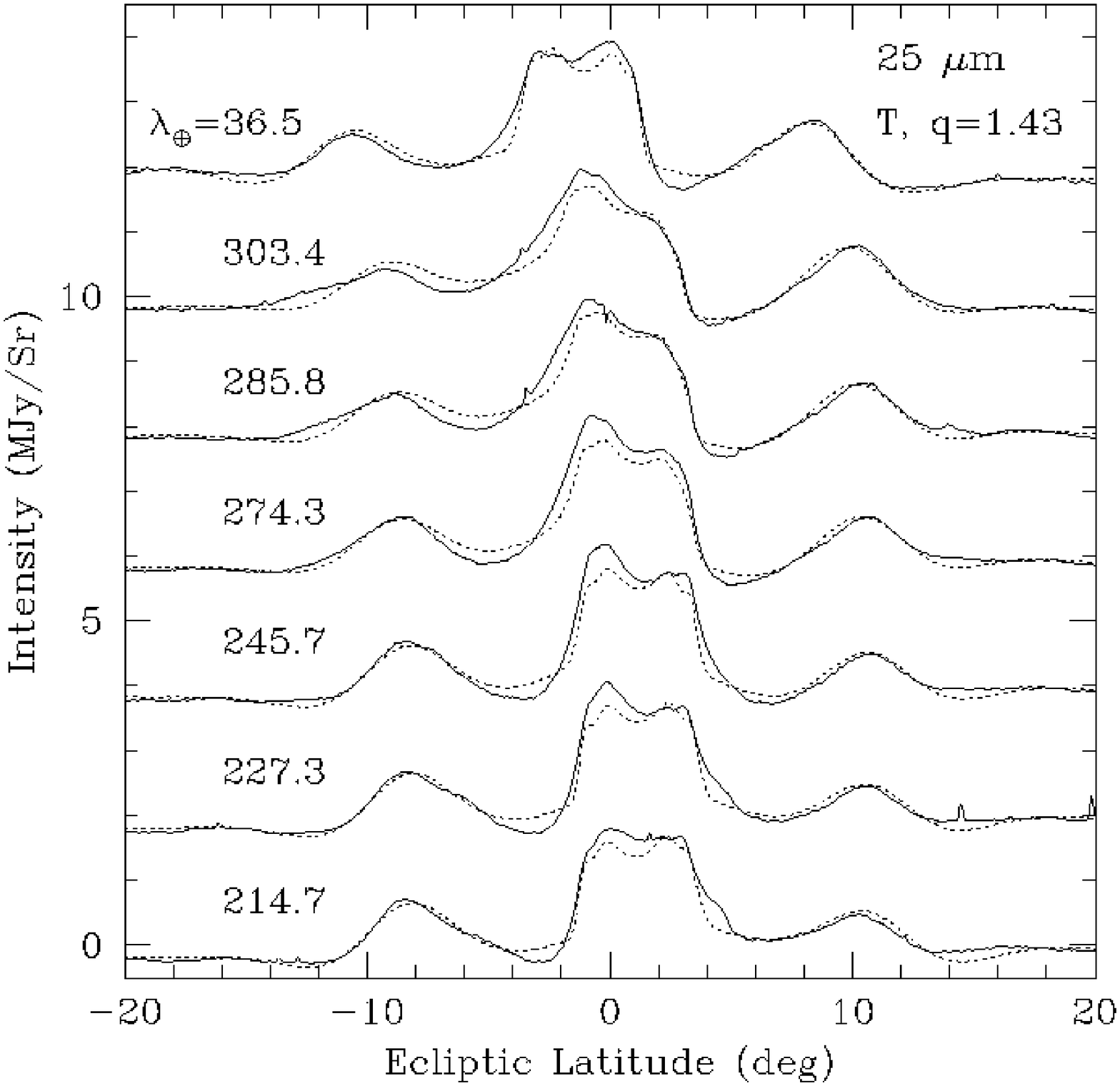,width=3in} \\
 \end{tabular}
 \end{center}
 \caption{Best fit dust band model (dotted curves) to the 25 $\mu m$ IRAS 
observations in both the leading (top) and trailing (bottom) directions. This 
model has a size-frequency index $q$ of 1.43, indicating that large particles 
dominate the distribution.}
\end{figure}

\clearpage

\begin{figure}
 \begin{center}
 \begin{tabular}{l}
 \psfig{file=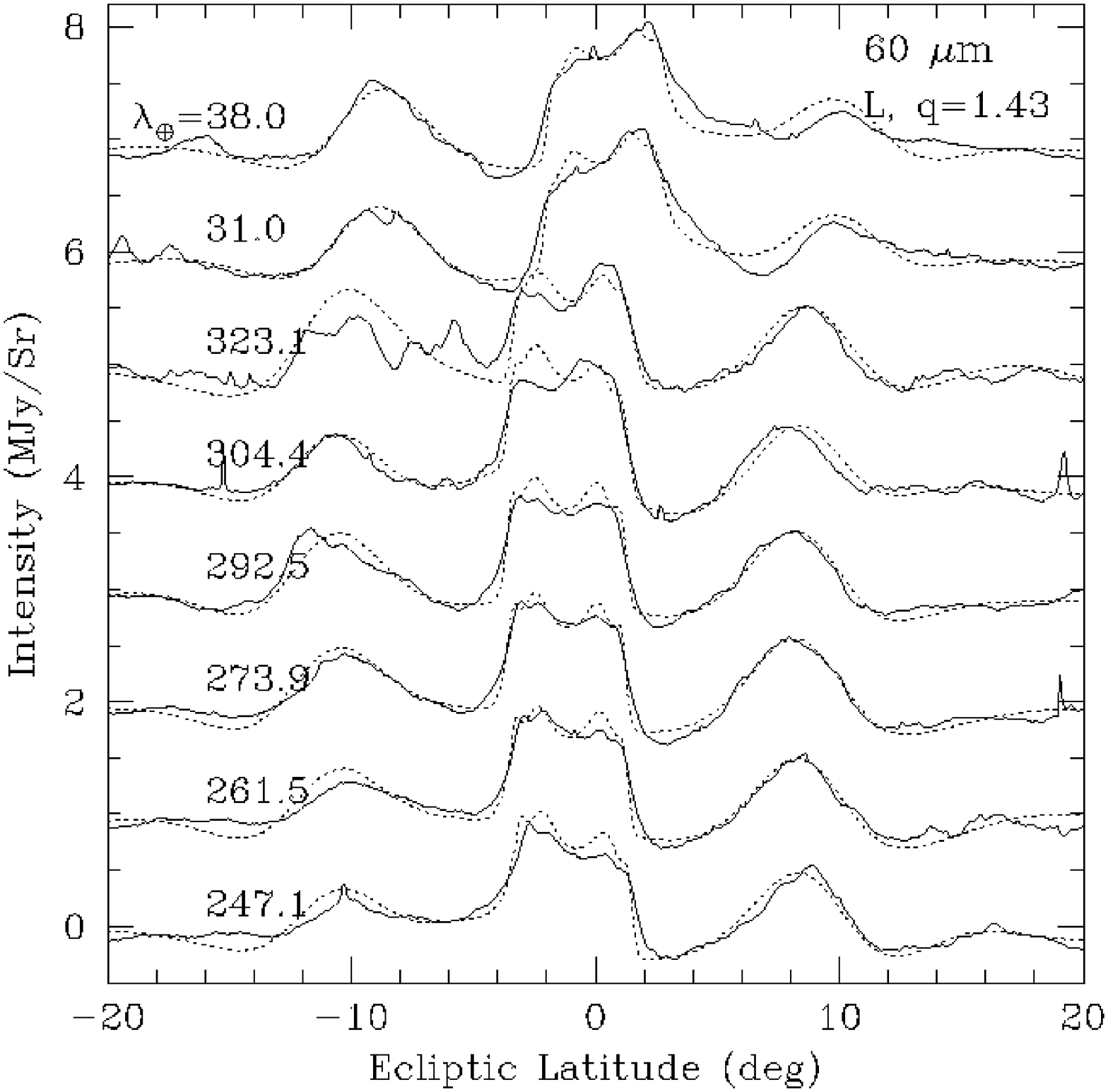,width=3in} \\
 \psfig{file=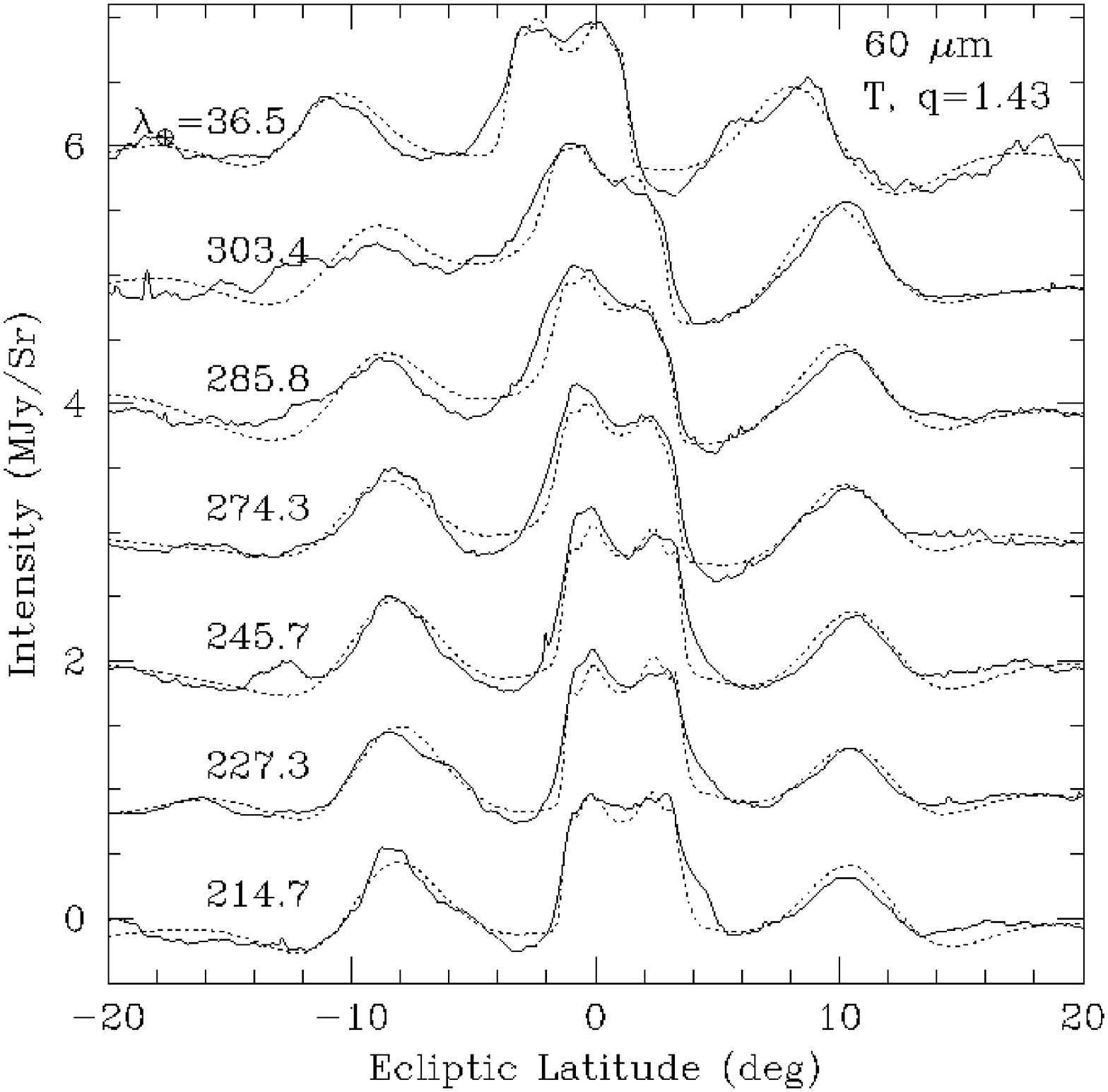,width=3in} \\
 \end{tabular}
 \end{center}
 \caption{Best fit dust band model (dotted curves) to the 60 $\mu m$ IRAS 
observations in both the leading (top) and trailing (bottom) directions. This 
model has a size-frequency index $q$ of 1.43, indicating that large particles 
dominate the distribution.}
\end{figure}

\clearpage

\begin{figure}[t]
 \centering
 \centerline{
 \psfig{file=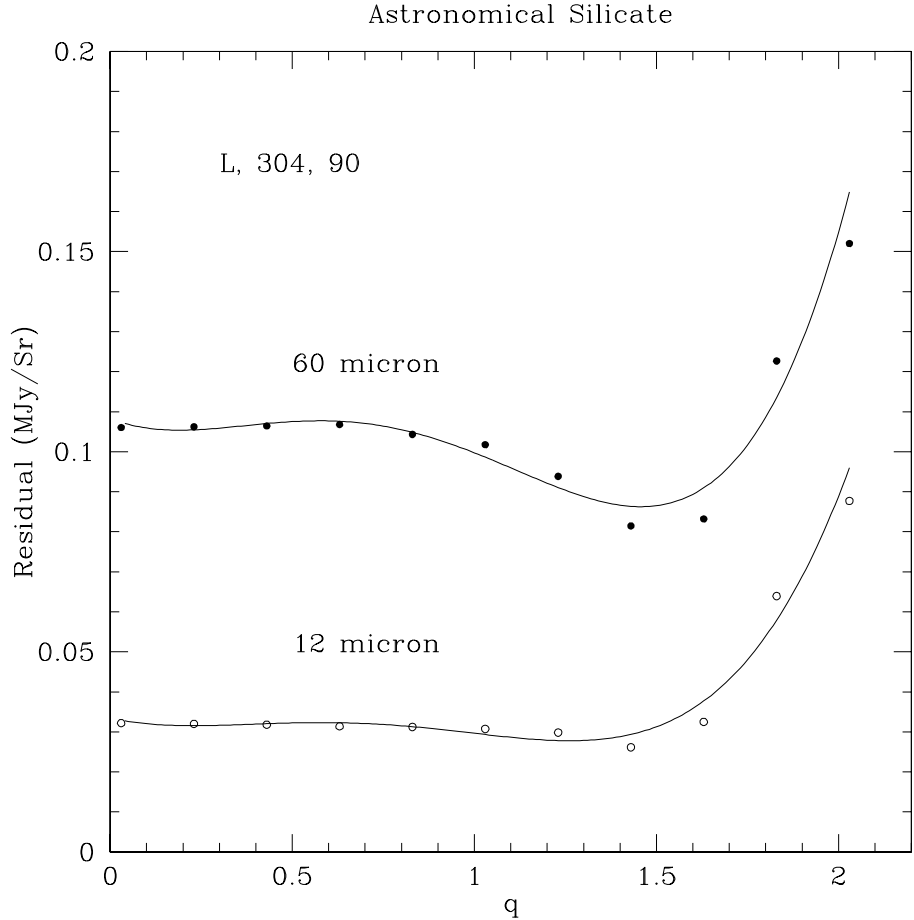,width=5in}
            }
 \caption{Goodness of fit of our models as a function of size-frequency index
$q$ for a single longitude of Earth. This has been obtained for the ten degree 
band by calculating the root mean square (observation - model) over two five 
degree wide latitude bins to cover the north and south bands for both the 12 
and 60 micron wavebands. When $q$ is too high, too many small particles are 
included in the model, and the amplitudes of the 12 $\mu m$ models are too 
large. In addition, too few large particles are included and the amplitudes of 
the 60 $\mu m $ models are too small. When $q$ is too low, too many large 
particles are included in the model. This leads to the distribution of forced 
inclinations in the model to be skewed too much towards the large end, and the 
model profiles are shifted in latitude with respect to the observations, 
degrading the fit. The particles are taken to be composed of astronomical
silicate.}
\end{figure}

\clearpage

\begin{figure}
 \begin{center}
 \begin{tabular}{l}
 \psfig{file=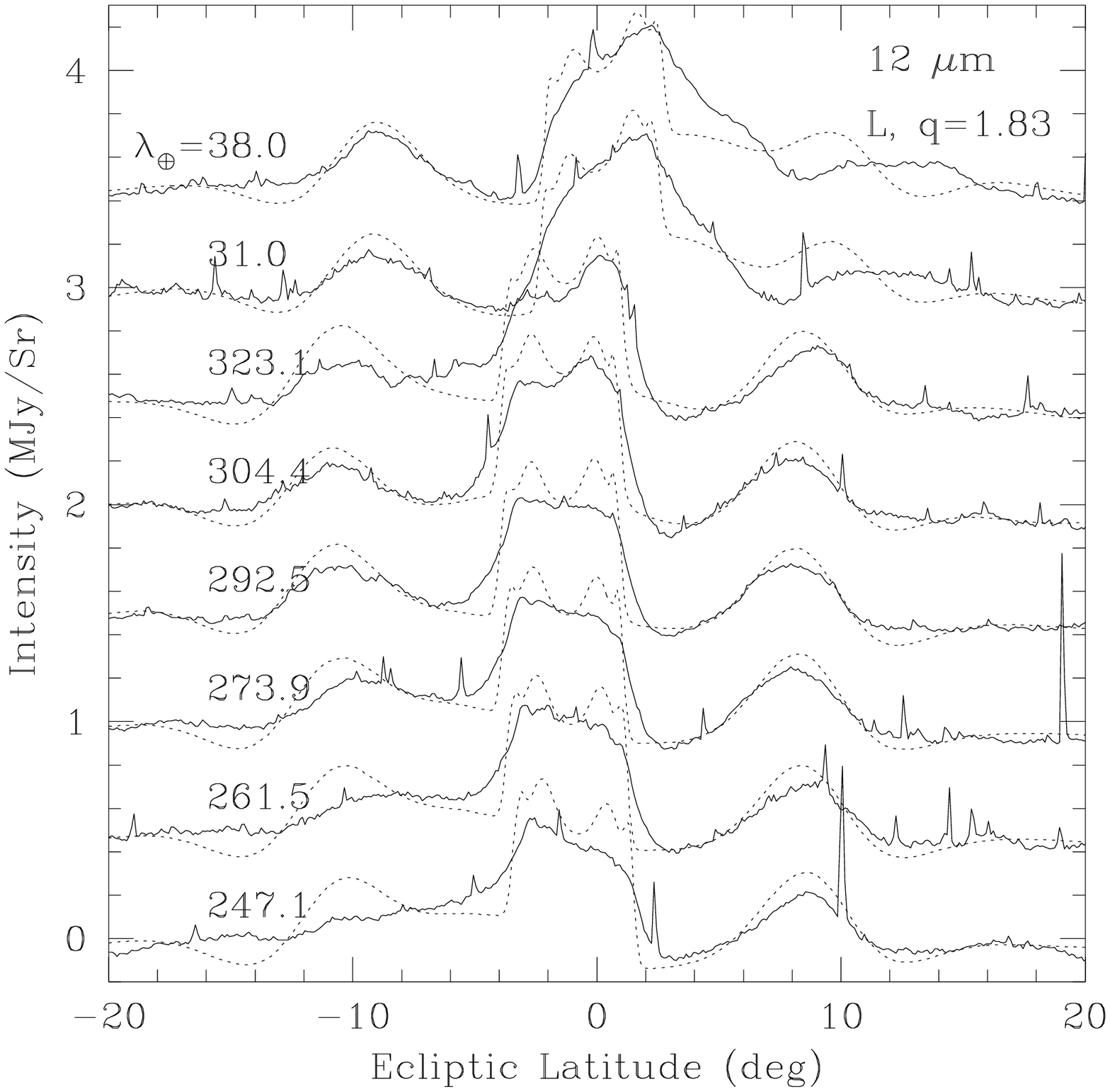,width=3in} \\
 \psfig{file=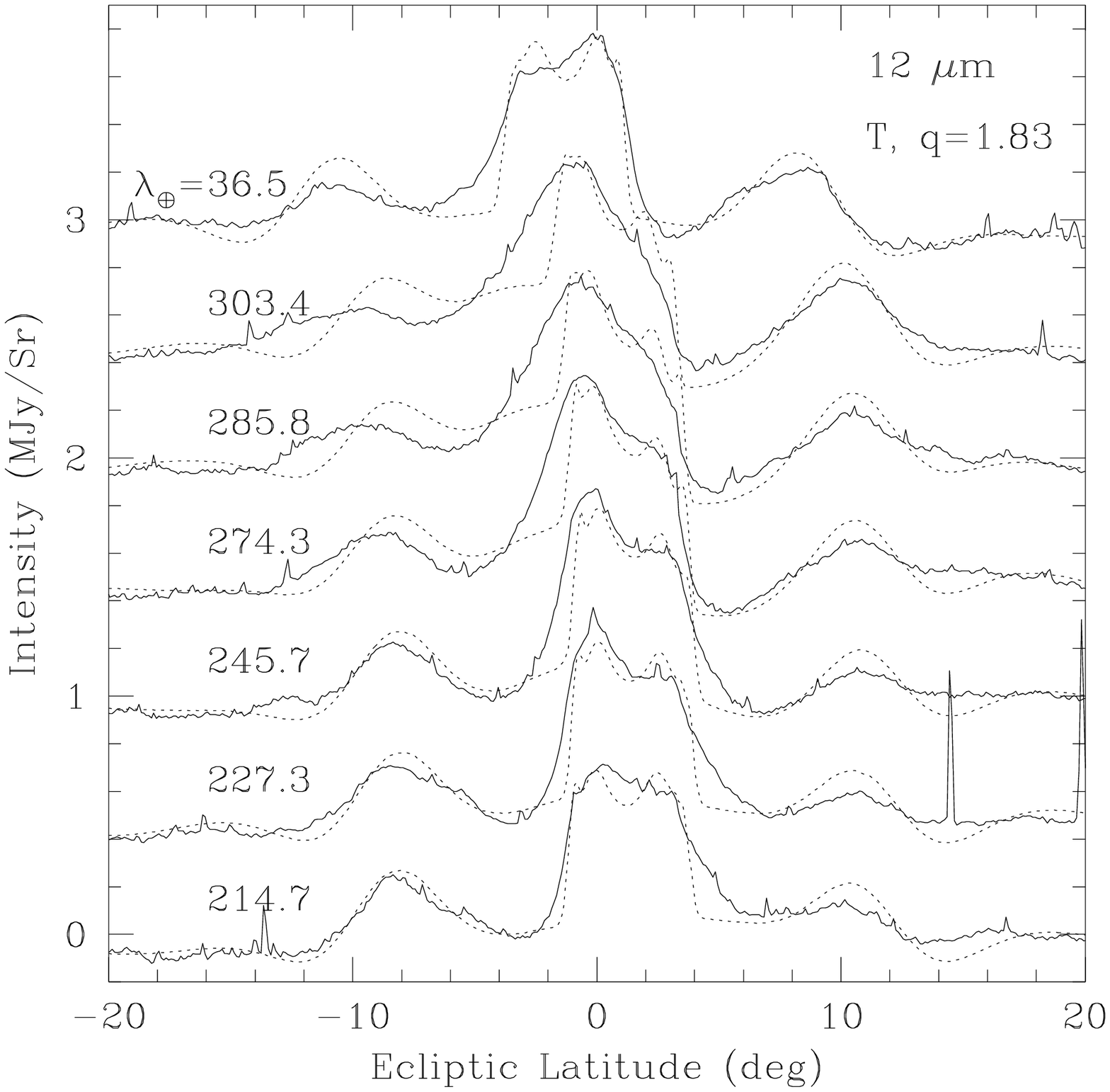,width=3in} \\
 \end{tabular}
 \end{center}
 \caption{Comparison of a q=1.83 dust band model (dotted curves) to the 12 
$\mu m$ IRAS observations in both the leading (top) and trailing (bottom)
directions. Too many small particles are included in the model, and the 
ampitudes of the 12 $\mu m$ models are too large.}
\end{figure}

\clearpage

\begin{figure}
 \begin{center}
 \begin{tabular}{l}
 \psfig{file=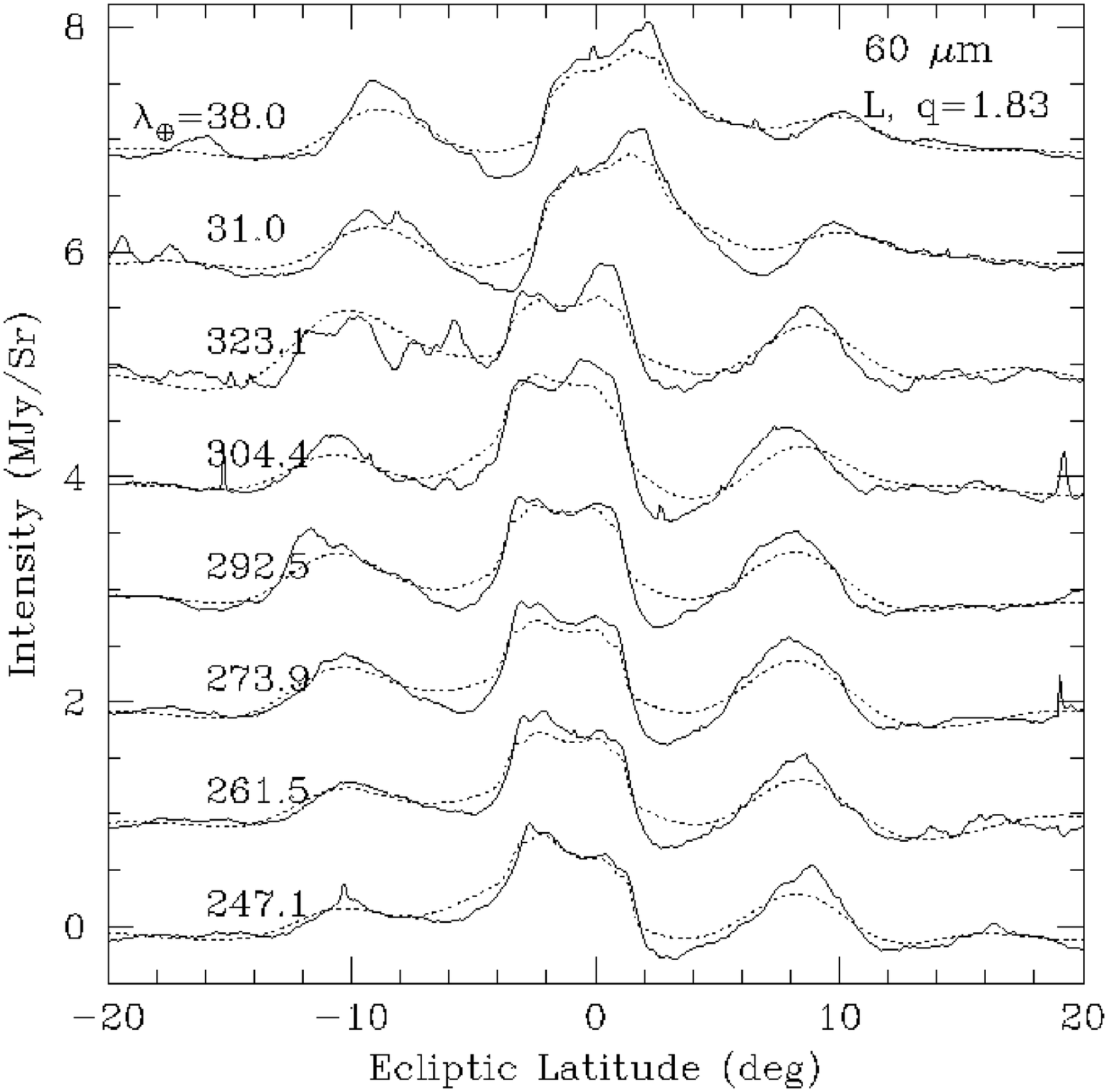,width=3in} \\
 \psfig{file=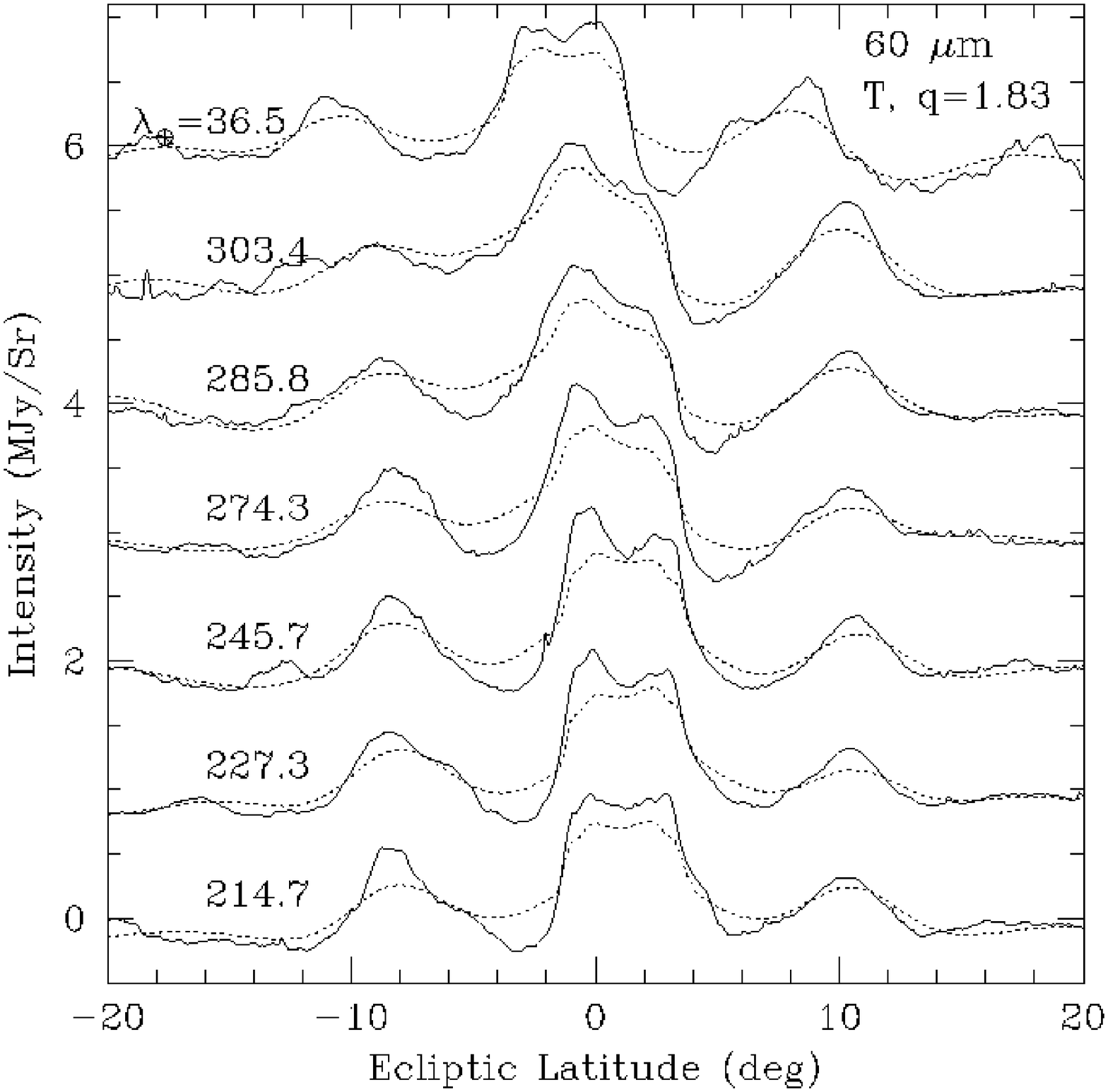,width=3in} \\
 \end{tabular}
 \end{center}
 \caption{Comparison of a q=1.83 dust band model (dotted curves) to the 60 
$\mu m$ IRAS observations in both the leading (top) and trailing (bottom)
(directions). Not enough large particles are included and the amplitudes of 
the 60 $\mu m $ models are too small.}
\end{figure}

\clearpage

\begin{figure}[t]
 \centering
 \centerline{
 \psfig{file=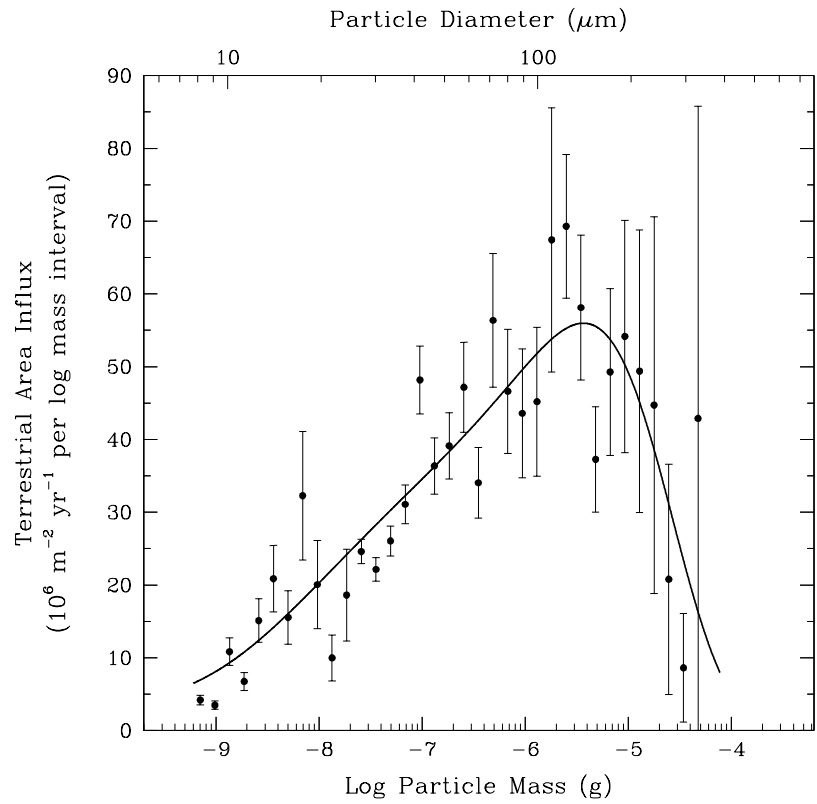,width=5in}
            }
 \caption{The terrestrial influx of zodiacal dust particles, as measured from 
the cratering record on the LDEF satellite. The slope of area against particle 
mass indicates a value for $q$, the size-frequency distribution index, of 
approximately 1.15.}
\end{figure}

\clearpage

\begin{figure}
 \begin{center}
 \begin{tabular}{l}
 \psfig{file=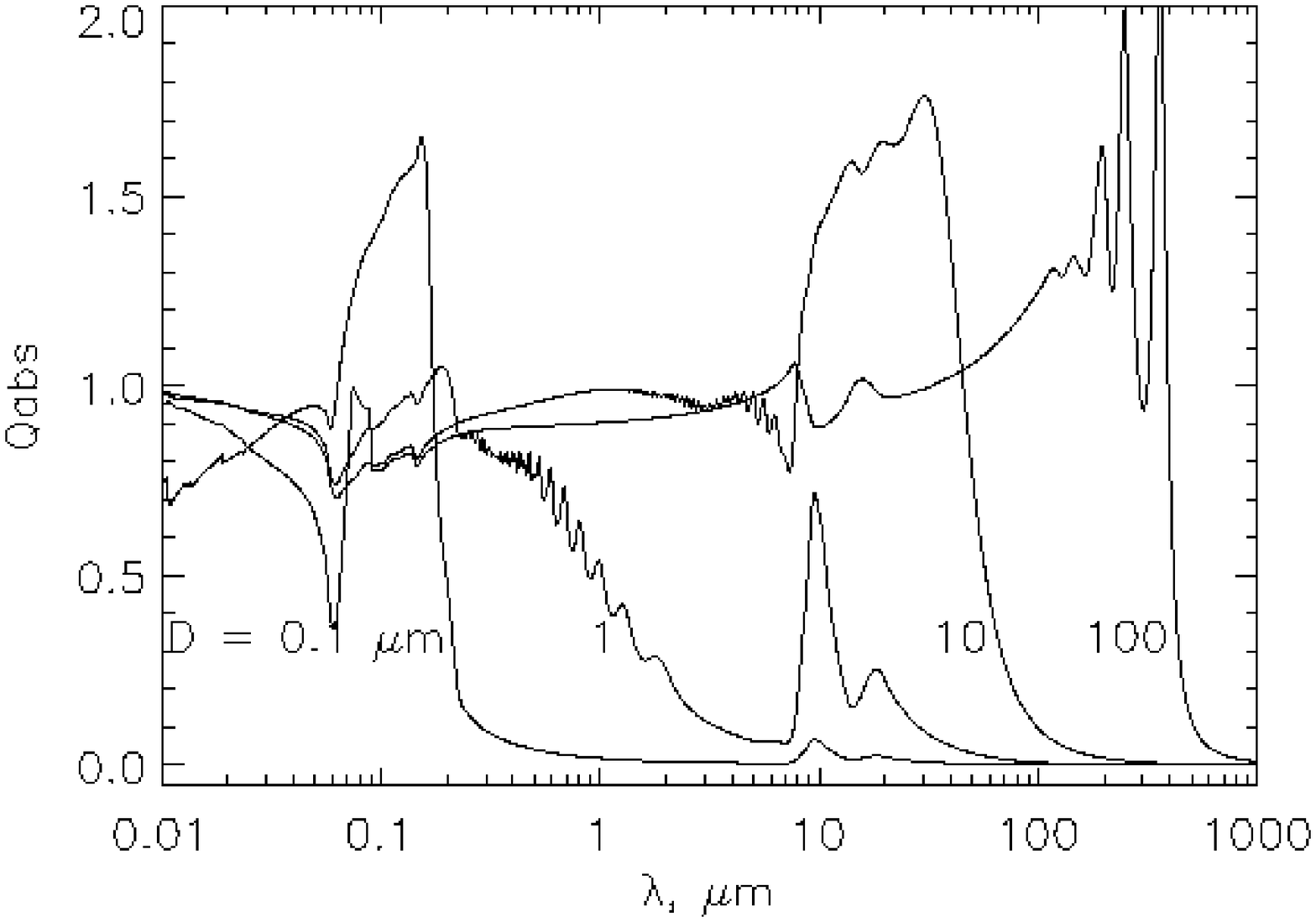,width=4in} \\
 \psfig{file=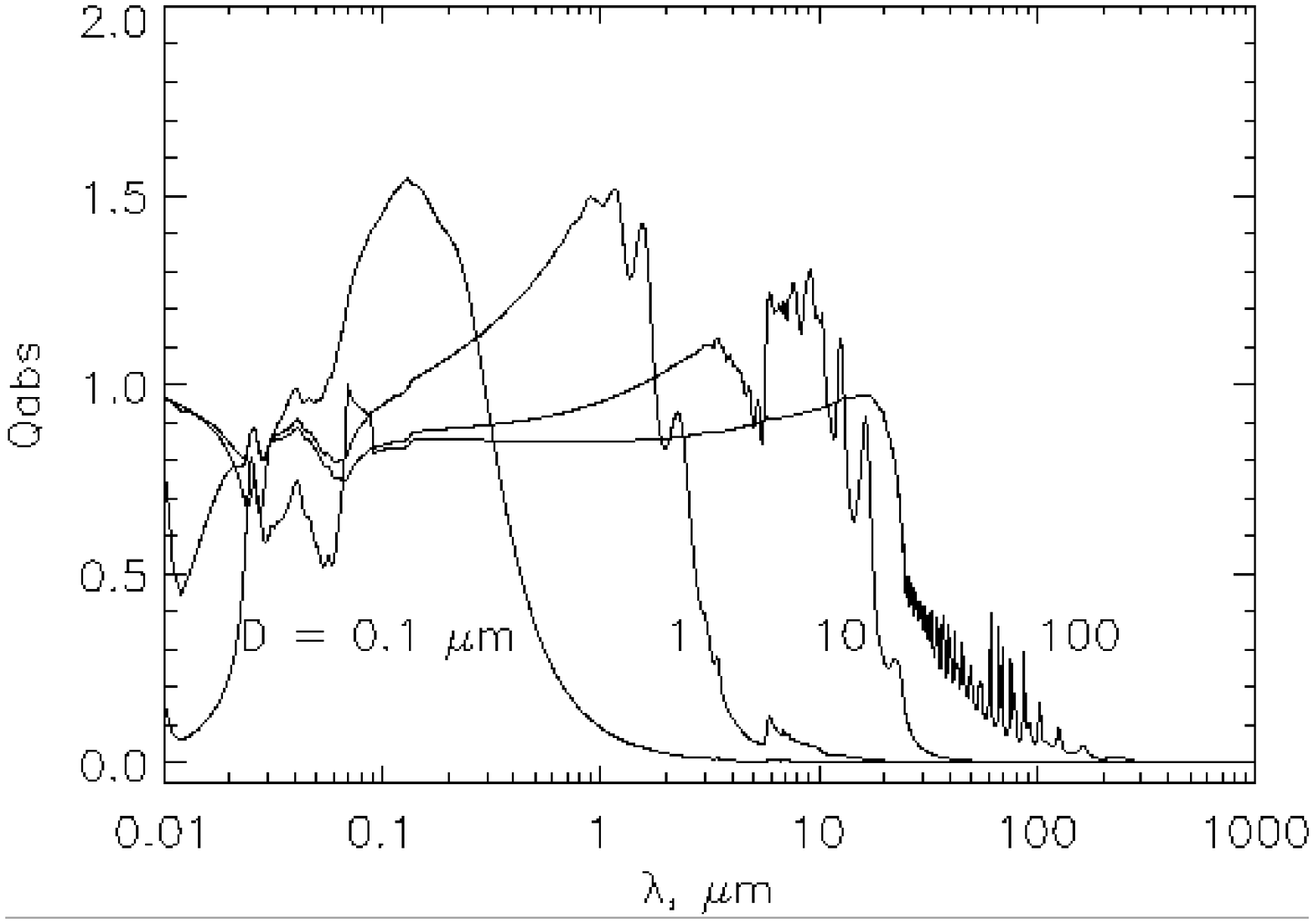,width=4in} \\
 \end{tabular}
 \end{center}
 \caption{Variation with wavelength and particle diameter of the absorption
 efficiencies of astronomical silicate (top) and organic refractory material 
 (bottom), calculated using Mie theory.
 }
\end{figure}

\clearpage

\begin{figure}[t]
 \centering
 \centerline{
 \psfig{file=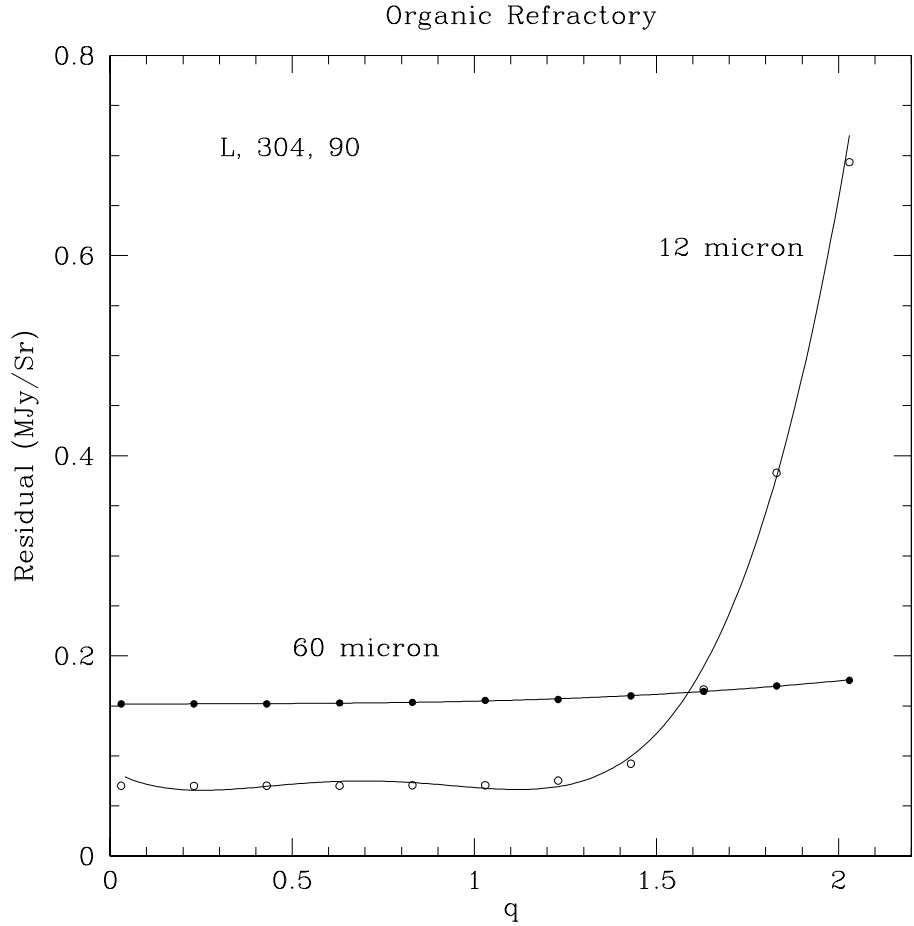,width=5in}
            }
 \caption{The results shown in Figure 15 are repeated for organic
 refractory particles, to investigate the dependence of the modeling on the
 dust particle properties. The 12 micron residuals strongly reinforce the 
 result obtained with astronomical silicate that a low size-frequency index $q$ 
 is required to match the observations. The low emissivity of organic 
 refactories at longer wavelengths means that the 60 micron residuals are of 
 little help in pinning down the size-frequency distribution index, but
 even here the residuals decrease as $q$ decreases.}
\end{figure}

\clearpage

\begin{figure}[t]
 \centering
 \centerline{
 \psfig{file=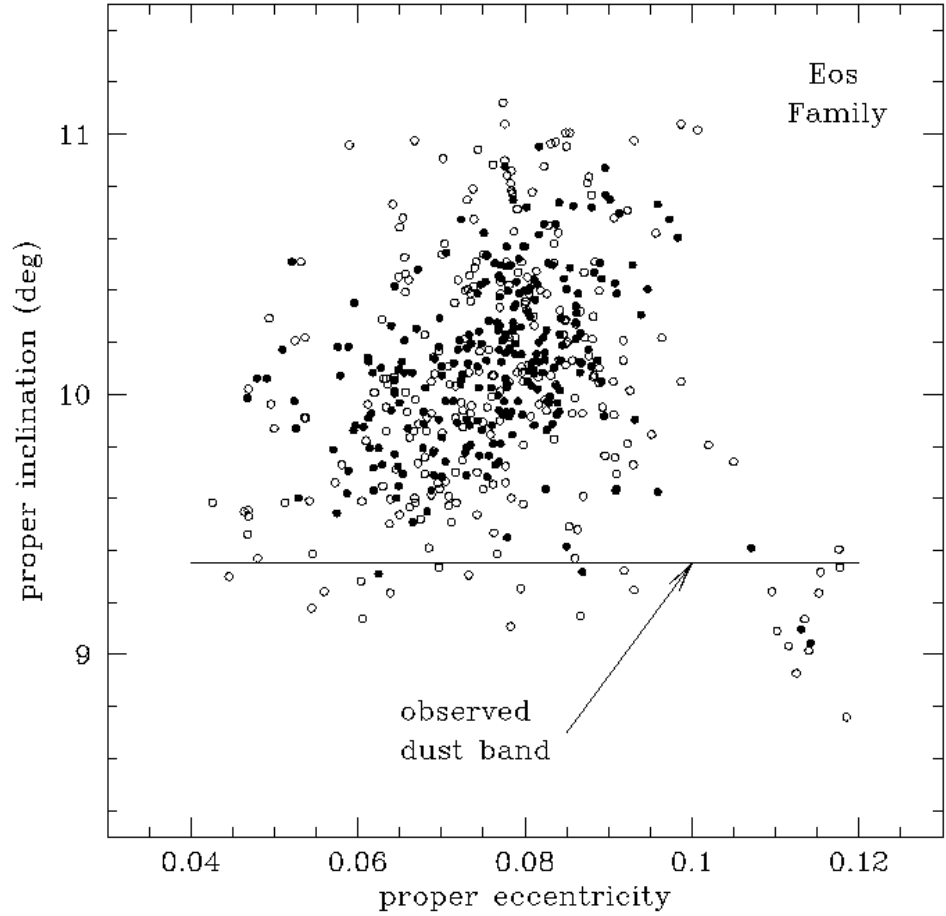,width=5in}
            }
 \caption{Eos family members, as determined by the hierarchical clustering 
method. Closed circles represent asteroids larger than 15km in diameter,
assuming an albedo of 0.12. The mean inclination of the model ten degree dust 
band material (9.35$^{\circ}$) is labeled. This suggests that either the ten 
degree band is produced from collisional activity near the inner edge of the 
family or that the inclinations of dust particle orbits from the Eos family 
as a whole no longer trace the inclinations of their parent bodies but have 
been degraded since their production.}
\end{figure}

\clearpage

\begin{figure}
 \begin{center}
 \begin{tabular}{ll}
 \psfig{file=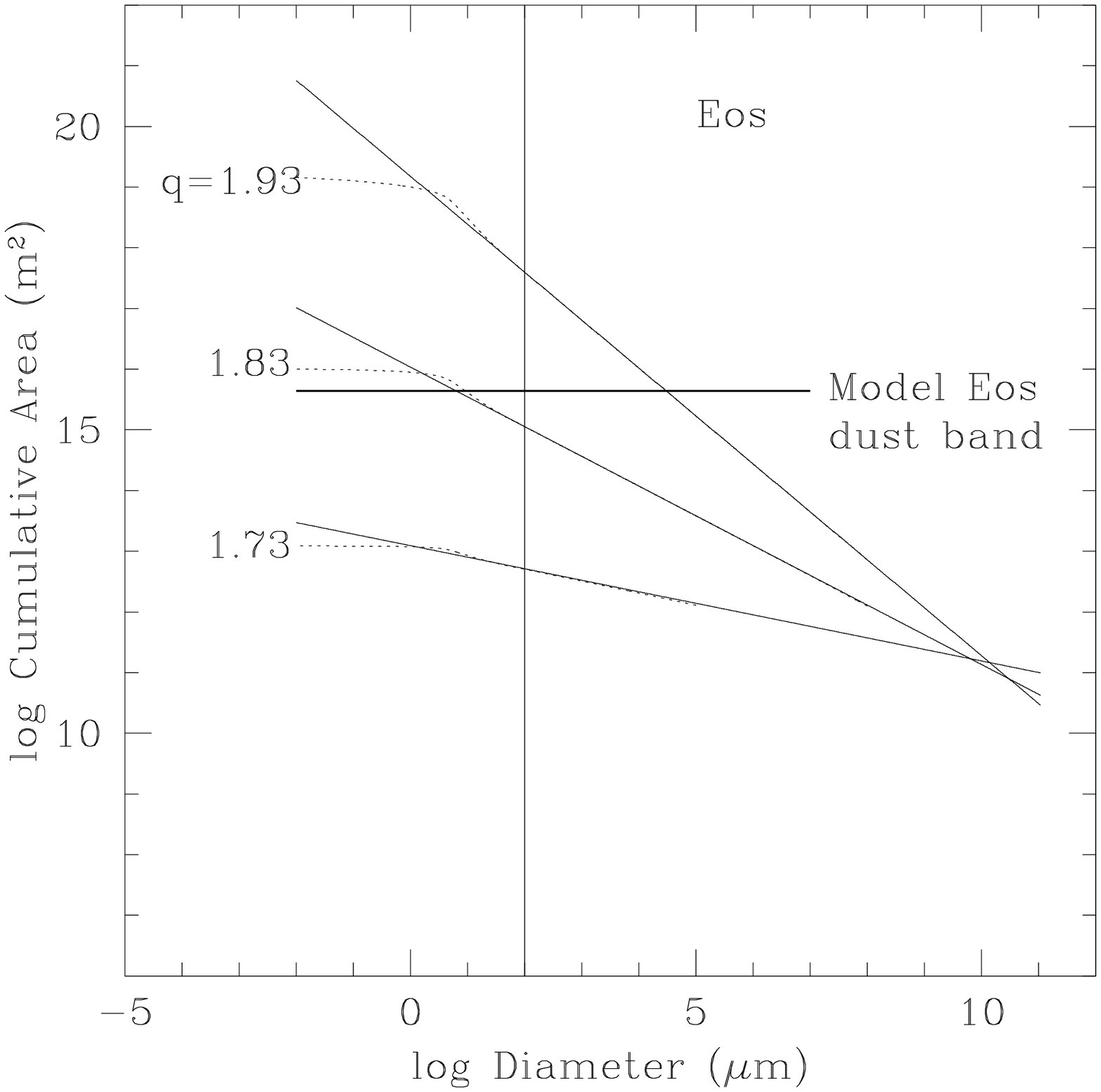,width=2.5in} &
 \psfig{file=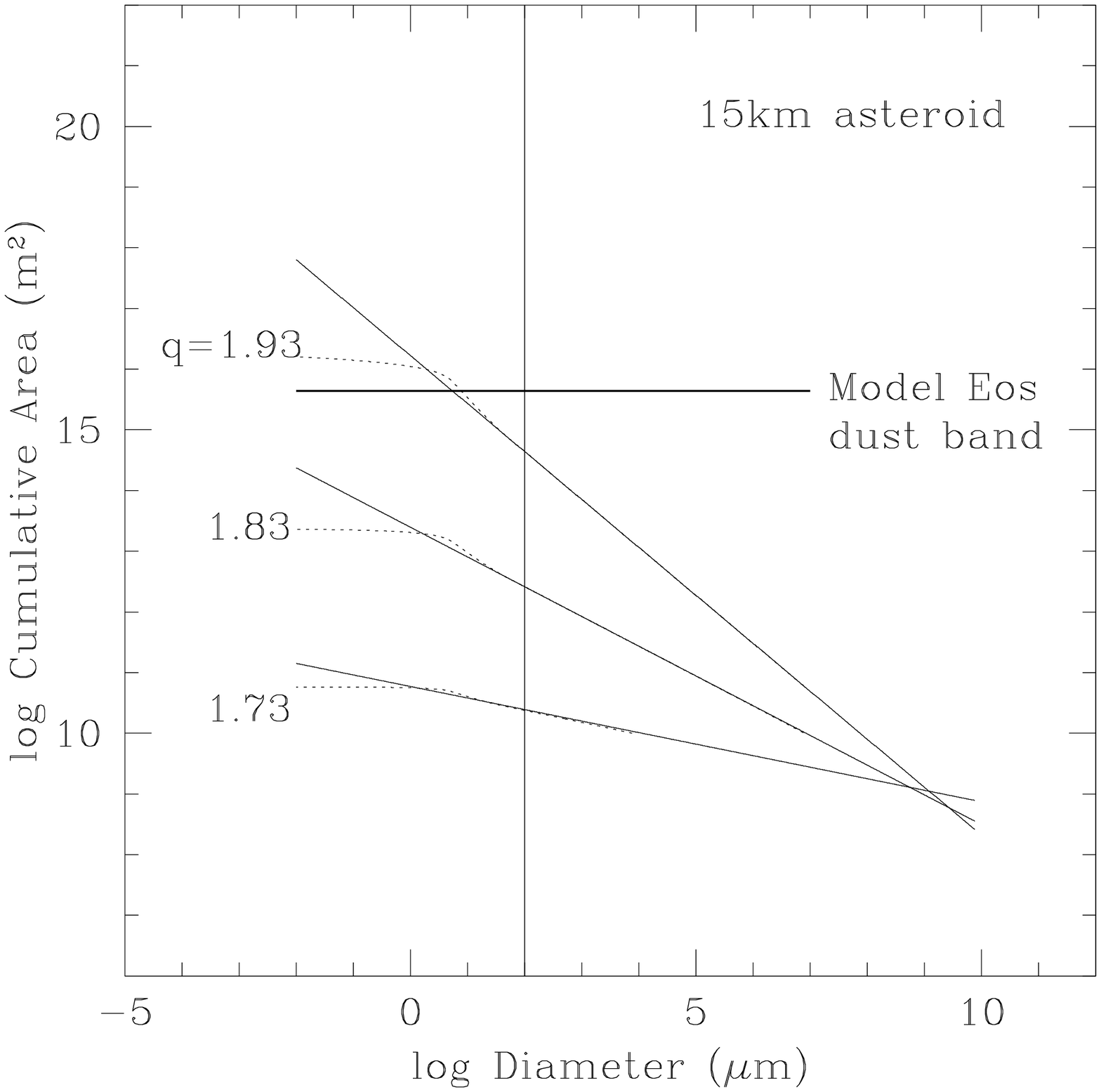,width=2.5in} \\
 \psfig{file=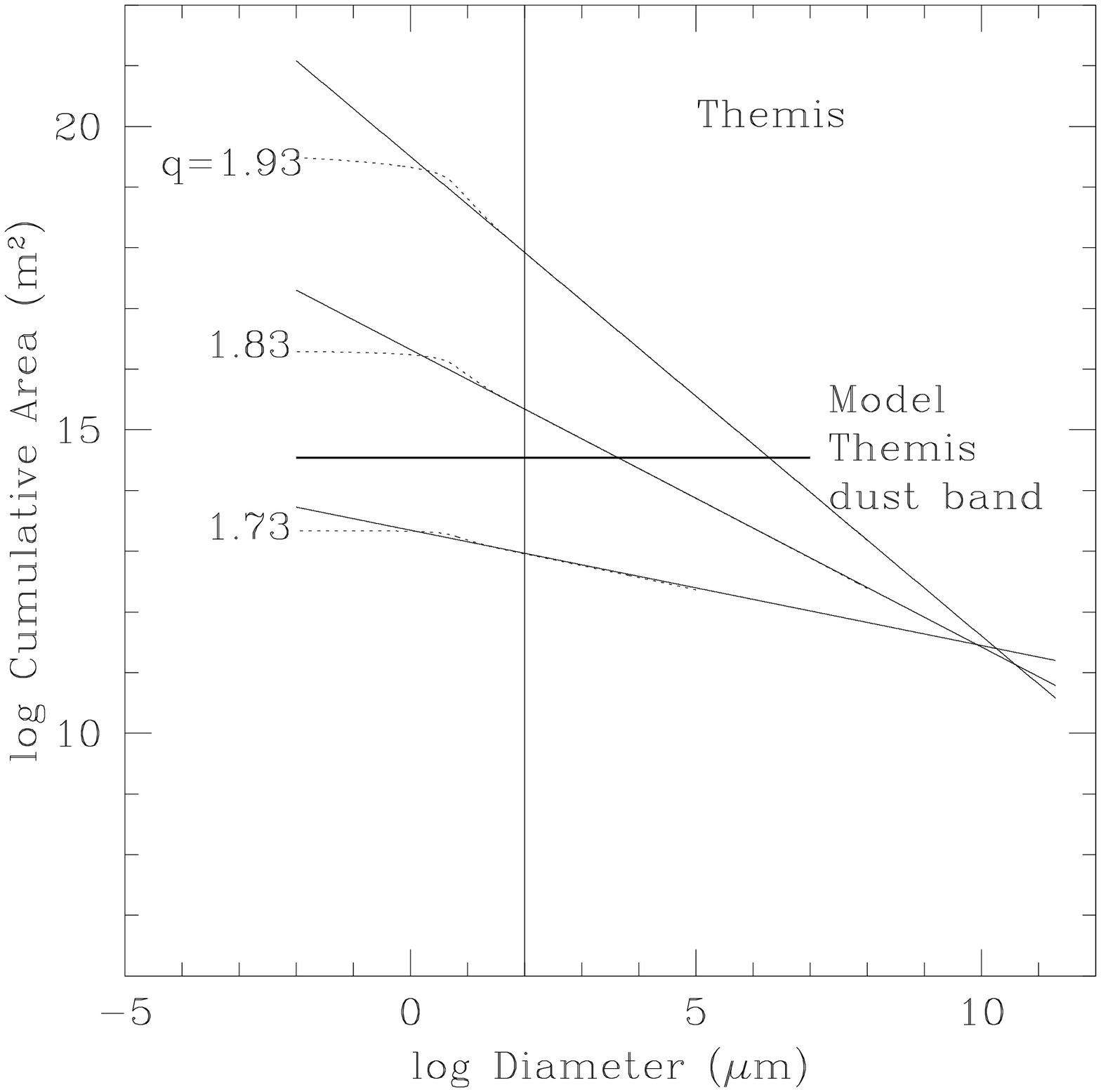,width=2.5in} &
 \psfig{file=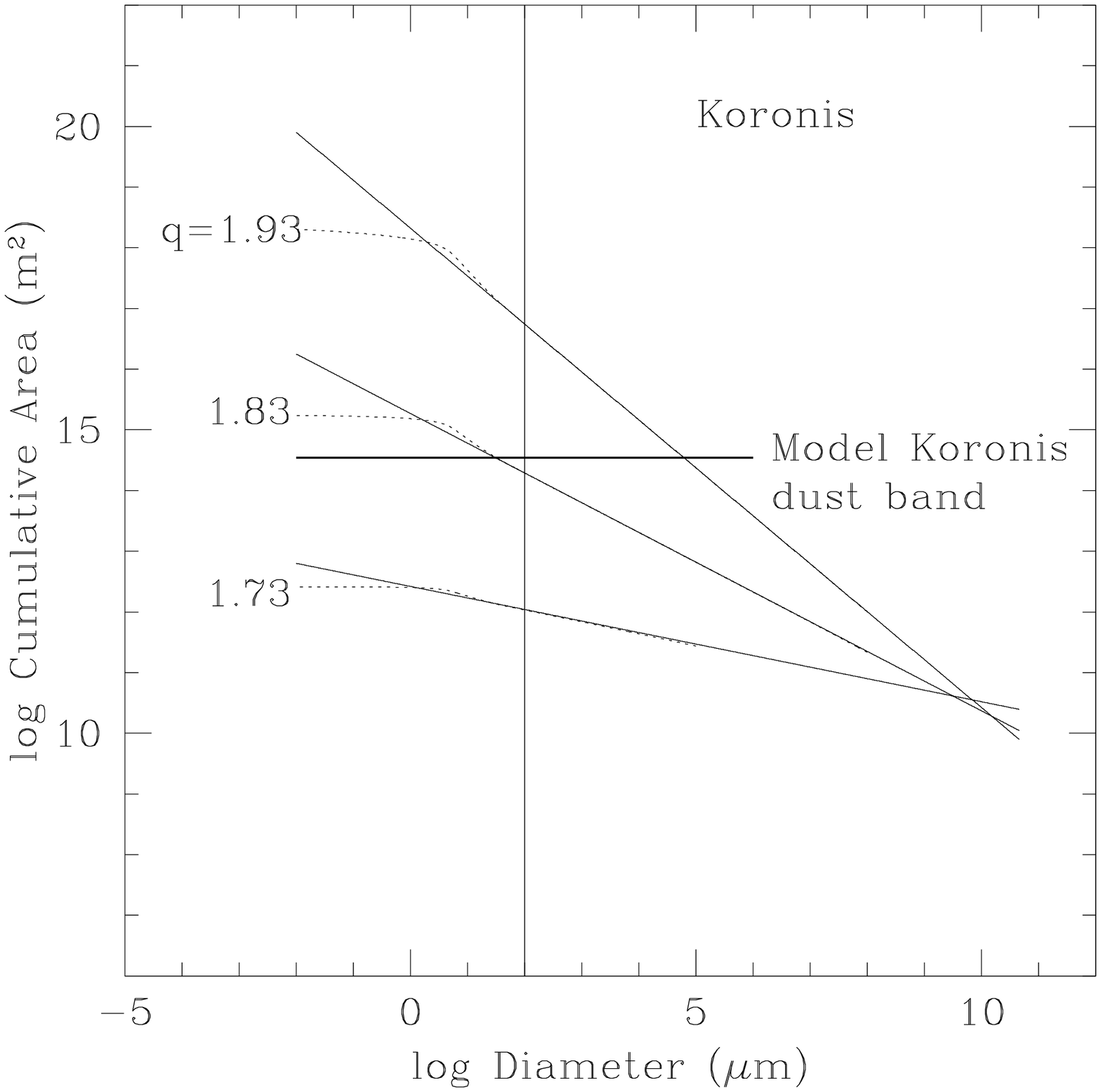,width=2.5in} \\
 \end{tabular}
 \end{center}
 \caption{The cumulative surface area as a function of different size-frequency 
distribution indices for the Eos, Themis and Koronis asteroid families and also
a single 15km diameter asteroid. The solid curves are for blackbody particles.
The dotted curves are for astronomical silicate particles; as the particle
size decreases below approximately 1 $\mu m$, $Q_{abs}$ decreases and the 
cumulative surface area begins to level off. The dust band modeling predicts
the amount of surface area associated with each family, these values are
labeled. The vertical lines represent the position of 100 $\mu m$ particles
in the size distribution.
For a single asteroid to be responsible for 
the ten degree dust band, the size-frequency index of the collisional debris 
would initally have needed to be extremely high to produce the surface 
area required to match the observations.}
\end{figure}

\clearpage

\begin{figure}[t]
 \centering
 \centerline{
 \psfig{file=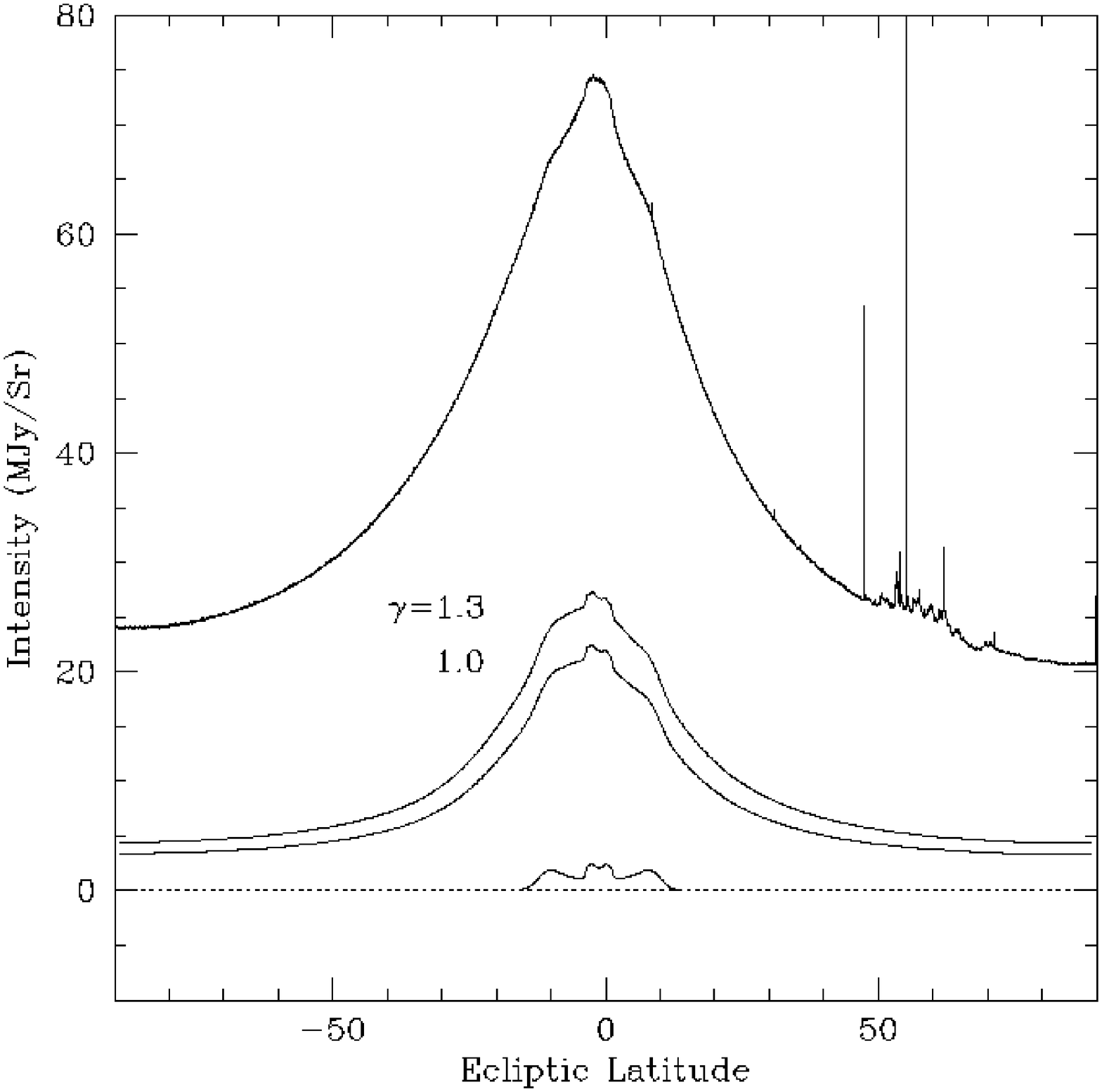,width=5in}
            }
 \caption{The contribution of dust band material to the zodiacal cloud.
The figure shows a comparison of the thermal emission obtained from the raw 
best fit dust band model to the corresponding IRAS profile in the 25 $\mu m$ 
waveband. The result is shown for inner Solar System distributions of
material corresponding to $\gamma$=1.0, as expected for a system evolved by
P-R drag, and $\gamma$=1.3 as predicted in parametric models of the 
zodiacal cloud, most recently Kelsall et al. (1998).  
The dust bands appear to contribute approximately 30\% to the total thermal 
emission. Also shown is the amplitude of the dust band material confined
to the main belt (exterior to 2 AU), which represents the component of the
dust band material isolated by the fast Fourier filter. This clearly shows
the extent to which the dust band contribution is underestimated if it is
assumed that the filtered dust band observations represent the entireity of 
the dust band component of the cloud.}
\end{figure}

\clearpage

\begin{figure}[t]
 \centering
 \centerline{
 \psfig{file=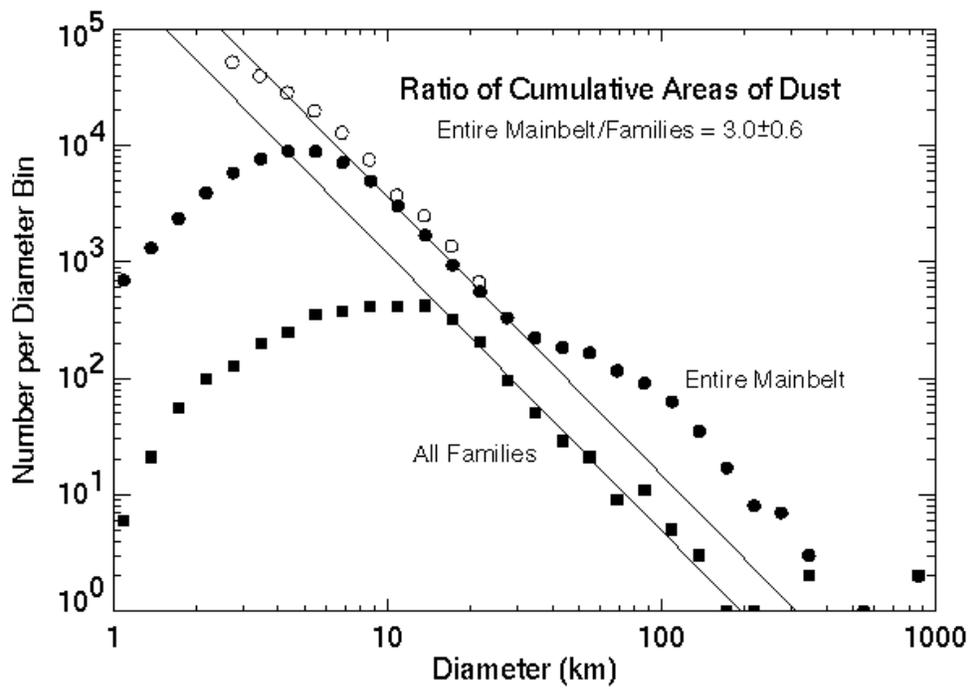,width=5in}
            }
 \caption{The ratio of areas of dust associated with the entire main belt 
 asteroid population and all families, obtained by extrapolating the observed 
 size  distributions of larger asteroids in both populations assuming a 
 collisional equilibrium power law size distribution.}
\end{figure}

\clearpage

\begin{figure}[t]
 \centering
 \centerline{
 \psfig{file=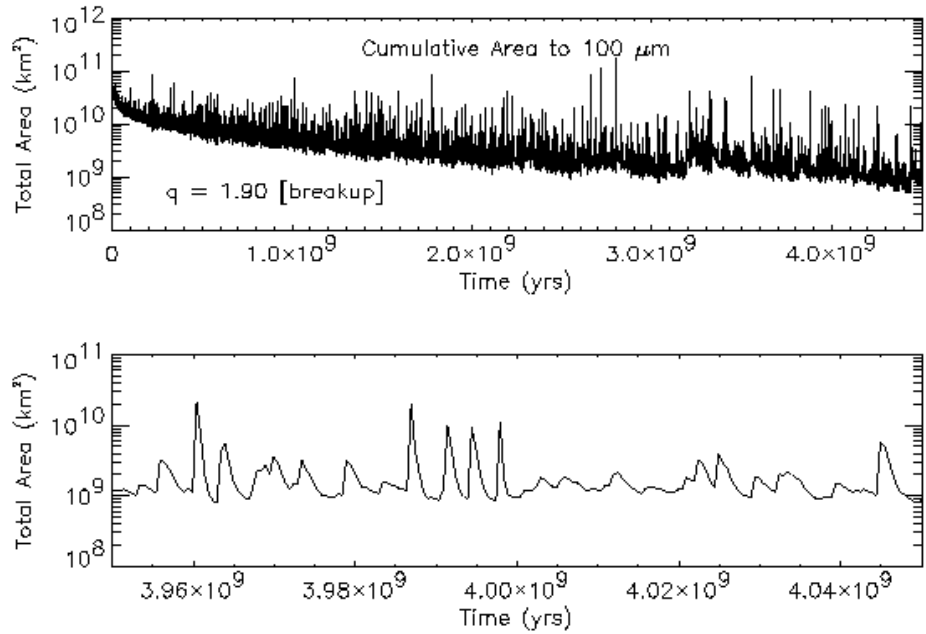,width=5in}
            }
 \caption{Variation of the total cross-sectional area associated
 with the main belt asteroid population over the age of the Solar System. 
 The initial mass is taken to be 
 approximately three times that of the current mass, and the fragmentation
 index $q=1.90$. The calculation
 is performed for particles from 100 $\mu m$ through the largest asteroidal 
 sizes, and describes the stochastic breakup of asteroidal fragments. The
 lower panel shows 100 million years of evolution 4 Gyr after the start of 
 the simulation.}
\end{figure}

\end{document}